# Surface Electromyography as a Natural Human-Machine Interface: A Review

Authors: Mingde Zheng*, Michael S. Crouch*, Michael S. Eggleston

Nokia Bell Laboratories. 600 Mountain Ave, New Providence, NJ, 07974 United States

* These authors contributed equally

**Abstract**

Surface electromyography (sEMG) is a non-invasive method of measuring neuromuscular potentials generated when the brain instructs the body to perform both fine and coarse locomotion. This technique has seen extensive investigation over the last two decades, with significant advances in both the hardware and signal processing methods used to collect and analyze sEMG signals. While early work focused mainly on medical applications, there has been growing interest in utilizing sEMG as a sensing modality to enable next-generation, high-bandwidth, and natural human-machine interfaces. In the first part of this review, we briefly overview the human skeletomuscular physiology that gives rise to sEMG signals followed by a review of developments in sEMG acquisition hardware. Special attention is paid towards the fidelity of these devices as well as form factor, as recent advances have pushed the limits of user comfort and high-bandwidth acquisition. In the second half of the article, we explore work quantifying the information content of natural human gestures and then review the various signal processing and machine learning methods developed to extract information in sEMG signals. Finally, we discuss the future outlook in this field, highlighting the key gaps in current methods to enable seamless natural interactions between humans and machines.

# Table of Contents

# 1 Introduction & Background

## 1.1 Rise of the Human-Machine Interface

Since the dawn of the 1$^{st}$ industrial revolution, we have sought effective modes of interaction with machines to help improve our efficiency and productivity. Early interaction with machines was dominated by simple mechanical actuators such as levers, ropes, and knobs which required significant human physicality. The advent of the Computer Age fundamentally transformed the way humans and machines interact, with the emphasis on physical interaction shifting to digital interaction. Blunt physical instruments were replaced by keyboard-controlled command line interfaces, mouse-navigated graphical user interfaces (GUI), and simple touch-based interfaces. Today, GUIs are ubiquitous in almost every sector of society, enabling flexibility of control parameters and input streams, while providing security and privacy features [1]–[3]. The increasing complexity and flexibility of our mechanized systems, however, has led to a corresponding increase in the complexity of these HMIs, leading to a strain on the cognitive workload of human workers [4]. Exacerbating this, we are now entering into the 4$^{th}$ industrial revolution, where the fusion of artificial intelligence (AI), robotics, Internet of Things (IoT), 3D printing, and other technologies are giving rise to a new age of cyber-physical connectivity, blurring the boundaries between digital, biological and the physical worlds [5], [6]. This explosion of linked devices and systems constantly increases the number of communication channels between humans and machines, challenging our mental capacities and requiring the development of ever more sophisticated HMI technology. In an attempt to increase the bandwidth of HMI without placing increasing burden on humans to learn artificial controls, there has been a move towards creating a more natural form of interactions with machines, known as a Natural User Interface (NUI) [7]–[9].

NUIs sense the user's body movements, voice inputs, and potentially even thoughts to create an experience where even a novice instantly feels like they have expert control [10]. Physically, NUIs rely on unobtrusive sensors embedded either on a person or in their immediate environment. Inertial measurement units (IMU) embedded in a wristband or glove, for example, have been demonstrated to track hand gestures via motion of the fingers and hand [11]–[14]. Video cameras are another common physical sensor employed [15], [16], with real-time video analytics techniques demonstrated that can interpret physical movements and body language [17]. IMUs and video analytics, however, are unable to fully capture the rich fine-grained and subtle motion of the human musculoskeletal system.

To bridge this gap, there is an emerging interest in using surface electromyography (sEMG) as a sensor for NUI [18]. sEMG is capable of measuring the neuromuscular potentials that the brain uses to communicate with the body. These myoelectric signals are generated as command instructions from the brain, delivering activation triggers to the skeletal muscles of the body for both fine and coarse human locomotion. This sensing capability does not require sophisticated hardware and it can be made either in a benchtop or portable scale; it does not require dedicated personnel to operate; and it can be constructed cost-effectively for a variety of applications. Most notably, it is non-invasive, only requiring adequate electrode contact with the skin to function. As the miniaturization of electronics continue to evolve alongside the development of more sophisticated machine learning techniques, sEMG has the potential to become a high-bandwidth, non-intrusive, and truly natural interface for future human-machine interaction [19]–[22]. The following sections overview the physiological basis of myoelectric signals, followed by a review of the major developments in the hardware and software research aimed towards enabling sEMG-based natural user-interfaces.

## 1.2 Myoelectric Physiology of The Human Body

When the brain instructs the body to move, it sends an electrical impulse signal down the spinal cord and through an intricate network of peripheral nerves to the targeted muscle. This neuronal signal is transduced into a muscular contraction by numerous neurons known as motor units, each consisting of a motor neuron (anterior horn cell), its axon, and all the individual muscle fibers it innervates (Figure 1). Upon arrival of the electrical impulse from the brain, the motor units quickly depolarize the cell membrane space of their respective axon terminals, leading to a propagating action potential wave that travels across the muscle fibers. Since an activation impulse from the brain can recruit multiple motor units, all the resultant motor unit action potentials (MUAPs) become superposed as their electrical signals radiate though the muscle. The resultant electrical signal can be sensed on the skin by the surface electrodes, giving the characteristics EMG signal [23].

Since motor units do not fire synchronously and contain both positive and negative components, the superposed EMG waveform is essentially random but with an overall amplitude correlated with the number, size and frequency of recruited motor units. MUAPs generated by individual fibers of a motor unit sum up by ways of amplitude reinforcement in both time and space, allowing varying speed and force of muscle activation. Temporal summation enables successive action potentials to gradually build up tension in a muscle fiber, whereas spatial summation summons additional muscle fibers nearby to be recruited, generating even greater tension. Smaller muscle motions with fewer recruited motor units therefore exhibit low amplitude EMG signals, while large muscle motions conversely result in high amplitude EMG signals [24].

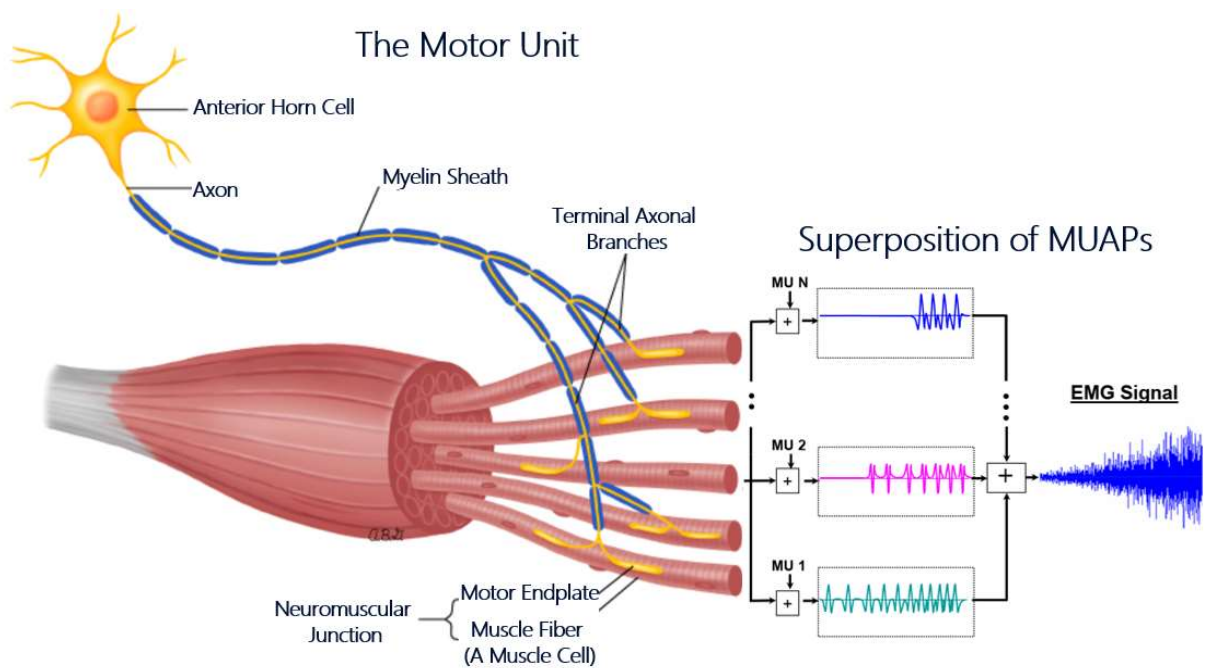

Figure 1: A single motor unit and the muscle fibers it innervates. When an anterior horn cell is activated, all muscle fibers depolarizes synchronously to generate a motor unit action potential (MUAP). Action potentials measurable by electrodes from all motor units superimpose to form the EMG signal.

## 1.3 Electromyography

Electrical activities of the skeletal muscles can be recorded by surface electrodes or needle electrodes. Needle electrodes are the clinical gold-standard method to evaluate individual motor units within a muscle. This approach, albeit invasive, provides detailed composition of the EMG signals and is advantageous for diagnosing medical conditions such as neuromuscular dystrophy or polymyositis [25]. EMG measurement through surface electrodes, on the other hand, lack the measurement specificity but are popularly embraced for being non-invasive. The detection of EMG signals through adhesive electrodes on the skin surface have been clinically beneficial in kinesiology studies of gait analysis and rehabilitation of prosthetic patients [26], as well as human-machine interface applications such as the control of robots and drones [27].

Universal protocols for EMG measurements are hard to define due to the diversity of applications and hardware configurations. However, general guidelines can be observed and have been predominantly documented in academic textbooks with minor cross-reference variations. It is generally accepted that prior to filtering and processing the superposed MUAPs, sEMG sensors should produce a raw signal ideally on a low-noise baseline, as seen in the example of Figure 2a where three biceps brachii contractions were executed with a rest interval in between. Averaged baseline noise level is a good assessment of overall EMG signal quality and should not exceed 3-5 microvolts in a high-fidelity system [28]. To obtain a high signal-to-noise (SNR) ratio, sufficient electrical contact with the skin must be achieved. "Wet" adhesive electrodes incorporating silver and silver-chloride (Ag/AgCl) metal is commonly recognized as the gold-standard (Figure 2b). Dry electrodes made of non-liquid based conductive materials such as stainless steel have been explored for enhanced user comfort that is free from adhesives and gel residues. However, they are prone to interfacial slippage, which causes unpredictable changes to electrode position and contact resistance, increasing noise and it is the main challenge to universal adoption [28].

The general analog process of EMG signal conditioning is illustrated by the flow chart in Figure 2c. Bipolar electrodes coupled with a differential amplifier is the most commonly employed measurement arrangement [29]. With this technique, a pair of differential electrodes are placed along the length of muscle fiber, and a third reference electrode on an electrically neutral site. In contrast, monopolar recording consists of a single recording electrode and a reference electrode. Bipolar arrangement is typically more advantageous since it offers common-mode electrical noise rejection and therefore higher

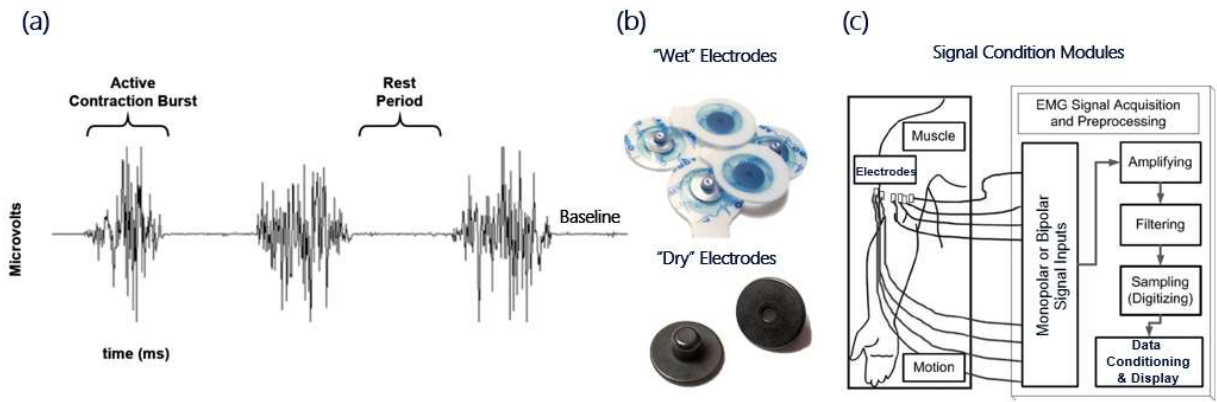

Figure 2: a) An example of raw EMG signal recording of 3 contractions bursts of the Biceps Brachii. b) Two types of electrodes commonly used, wet and dry. c) Signal conditioning modules that amplify, filter, sample and display EMG for analysis.

SNR [30], [31]. The un-amplified EMG signal amplitudes measured on the skin are only a few microvolts to millivolts, therefore the signal is always amplified by a factor of at least 500 to 1000 to match input voltage ranges of commercially available analog-to-digital converters (ADCs). Input impedance of the amplifier is typically in the range of 1 – 10 Megaohms [32]. The frequency range of the EMG signal is a frequently adjusted parameter, typically performed by an analog bandpass filter to capture either the full range or selective region of the signal spectrum. sEMG signals typically have a frequency content ranging between 10 – 500 Hz, with dominant frequency power from 20 – 150 Hz depending on the skeletal muscle being measured. Before the EMG signal can be analyzed on a computer, it must be converted from an analog voltage to a digital format. This conversion process is typically performed through a 12-bit ADC with a dynamic range capable of capturing the full signal from noise floor to peak EMG amplitude. The ADC sampling rate must be sufficiently high to capture the full bandwidth of EMG frequencies, which according to the Nyquist Sampling Theorem is double the highest desirable frequency component. Literature reports vary widely on the selection of sampling frequency depending on the application, but typically range from 200 Hz to 2 kHz and beyond [33], [34].

## 2  Myoelectric Devices & Applications

Over several decades of development, sEMG has transformed from a medical-grade tool utilizing expensive laboratory equipment and burdensome wet electrodes towards a more miniaturized and user-friendly form. With a goal of minimizing the intrusiveness of sEMG so that it can blend naturally into the background, device research has focused on creating consumer-grade wearable devices [35]–[41], fabric-embedded systems [42]–[49], and recently, printable "tattoos" [50]–[53] that conform to the body without restricting free movement. In the following section, we will review this hardware evolution and discuss potential future advances in sEMG aimed towards creating a natural user interface.

### 2.1  Traditional sEMG Systems

Early use of sEMG required a simple hardware assembly: a few pairs of wet electrodes, a signal conditioner and an analog-to-digital converter (Figure 3a). Early demonstrations often focused on simple hand gesture detection, such as binary "on" or "off" commands to interact with remote-controlled toys. For example, Kim et al. [54] showed one pair of wet electrodes placed on the forearm anterior muscles to be sufficient in moving a toy car in the forward and backward direction . In a similar fashion, Li et al. [55] used five pairs of wet electrodes and benchtop electronics to control an aerial drone in four directions with four coarse but unique hand gestures (fist, wrist flex, wrist up, and ring finger flex) (Figure 3b). More complex hand and finger gestures detection have been widely demonstrated, while exploring different forearm electrode placements for optimal recognition accuracy [56]–[58].

Commercially available dry electrodes specifically designed for sEMG applications have also been investigated to carry out gesture recognition with greater user comfort [59], [60]. Because intimate skin contact is the key to acquire high quality sEMG signal, a sustained amount of external pressure must be exerted to prevent interfacial slippage during hand movements. As a result, this is often reflected by using tighter, fitted bands [61] or pressure-adjustable wraps [62].

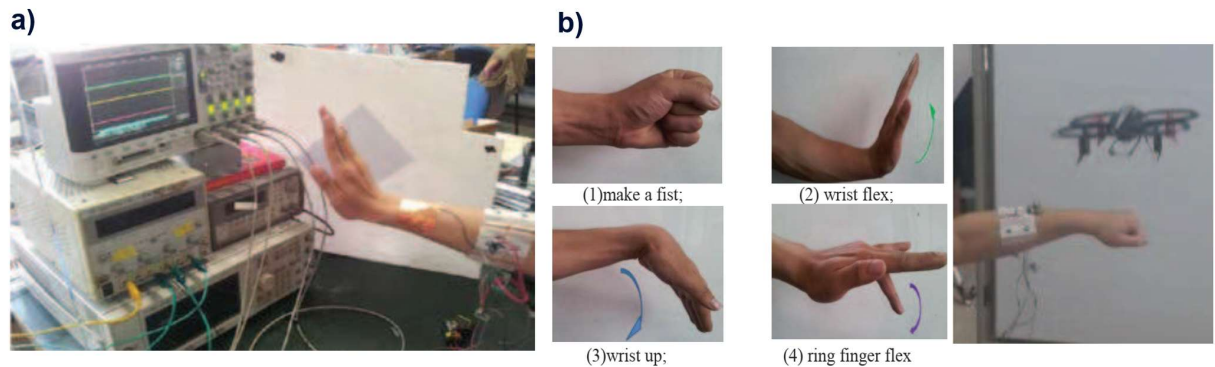

Figure 3: a) Typical sEMG benchtop hardware setup to obtain gesture differentiating signals. (b) Control of aerial drone with 4 hand gestures enabled by sEMG benchtop system. Figures reprinted with permission from [55].

Prosthetic control was among the first practical applications of sEMG as an HMI. Early benchtop sEMG systems demonstrated the control of remote robotics, such as the tele-operation of a prosthetic arm by Shenoy et al. [63] and a sophisticated toy robot by DaSalla et al. [64]. This effort was further extended to enhance the physical capabilities of healthy subjects. For instance, Woczowski et al. [65], Lenzi et al. [66], and Chen et al. [67], investigated sEMG's ability to actuate exoskeletons. More recent research focused on enabling amputees to control external assistive devices by placing wet electrodes on intact skin adjacent to the amputation site in an attempt to restore their ability to gesture [60], [68] or access computer software application such as computer-based multimedia player [69]. While these early works demonstrated many successes in sEMG for HMI, they were limited in accuracy and functionality, which motivated the field to experiment with increasingly larger number of channels, in the hopes of achieving a higher bandwidth interface between human and machine.

## 2.2 High Density sEMG Systems

High Density sEMG (HD-sEMG) became popular due to sEMG's versatility to easily scale-up the sensors from a few one-dimensional electrode arrays on targeted muscles to a dense, two-dimensional array coverage on large areas of the skin. It is a high-resolution myoelectric signal measurement system capable of measuring all major (e.g. Biceps Brachii on the upper arm) and minor (e.g. innervation of wrist extensors) biopotential signals of the skeletal muscles. Extending the measurement channels to a 2D grid having sufficiently small electrode size and interelectrode spacing allowed methods to be developed that could decompose sEMG interference patterns to analyze single MUs and their individual firing events [70]–[72]. Furthermore, it became possible to extract topological information from the muscles to construct spatial montages of MU activities that map and elucidate the functional properties of muscles [73], [74]. Through multi-channel ADC hardware with simultaneous data acquisition, the number of sEMG signal streams can be scaled as high as the ADC hardware and software can support. There is not a consensus on the minimal number of channels that constitutes a HD-sEMG, though common acquisition systems employ anywhere between 32 and 128 sensor channels in either monopolar or bipolar configuration.

HD-sEMG maps showing intensity and spatial distribution of muscle activation are particularly useful in differentiating between large numbers of gestures and quantifying the strength of muscle exertion. Rojas-Martínez et al. were amongst the first researchers to take advantage of HD-sEMG to create topological images of muscle activation patterns and their association with movement. They defined and constructed time-averaged 2D intensity maps of myoelectric signals, where each pixel in the map represents the root mean square value in a time window of a sensing channel [33]. They showed that the averaged HD-sEMG maps built from different types of tasks over a group of subjects can enhance the usability and control of sEMG, especially in HMI applications. These images were created from three bands of 2D electrode arrays that were custom-made for five muscle groups in the upper arm and forearm of a subject, totaling 350 monopolar sensing channels (Figure 4a). They demonstrated that not only can these averaged HD-sEMG maps (Figure 4b) distinguish between four tasks at the elbow joint (forearm pronation, supination, elbow flexion and extension) and their exertion strength, but the availability of large number of sensing channels also enabled the development of automatic algorithms to detect low-quality signals and active muscle zones.

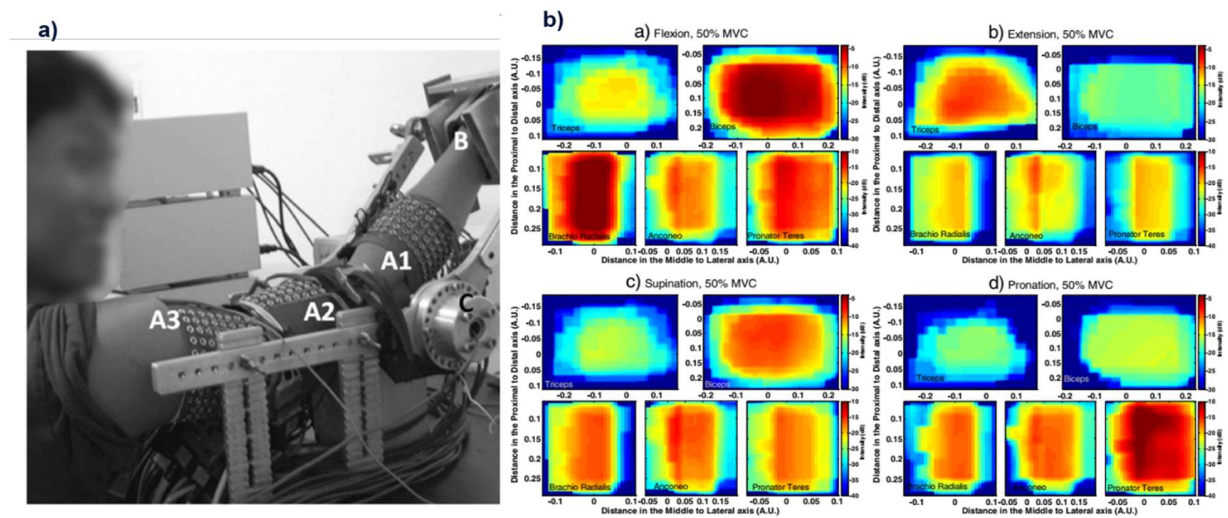

Figure 4: a) High density EMG array electrodes covering major muscles of both the upper and lower arm. b) Average HD-EMG maps across multiple subjects for four gestural tasks: Flexion (top left), Extension (top right), Supination (bottom left), and Pronation (bottom right). Figures reprinted from [33] (CC BY 4.0).

Furthering this work to develop a more fluid and natural HMI, Geng et al. showed that an instantaneous map of HD-sEMG can describe the physiological processes underlying muscle activities at a specific time, with the resolution defined by the electrode array [75]. In order to determine the existence of reproducible patterns across trials of the same gesture and differentiability of hand gestures among a group of individuals in these instantaneous HD-sEMG measurements, Geng et al. employed a convolutional neural network-based image classifier to verify that spatial patterns exist and that muscle activation signals of gestures from three databases can be subsequently recognized with accuracy up to 89.3%. This work provided a great starting point for enabling natural HMIs through HD-sEMG gesture recognition with the intention of reducing observational latency. As we will discuss in later sections, low latency algorithms with computational footprints that can be embedded in a wearable device are key requirements to fully exploit fine-grained human motion as a NUI.

In addition to increased repeatability, HD-sEMG is also able to detect and recognize subtle hand motions and muscle fatigues that prove difficult for simpler sEMG systems. Amma et al. [76] have demonstrated that HD-sEMG permits the detection of fine-grained finger motion by using a printed electrode array enclosing 168 differential sensing channels on the forearm to discriminate 27 finger gestures at 90% average intra-session accuracy. Their analysis indicated that increasing the number of electrodes would lead to greater performance gains. This work is significant in that low-amplitude sEMG signals from soft motions or low-strength exertions are part of the diverse sets of natural human body movements that could be used to interface with machines and devices, their reliable detection and recognition could greatly extend HMI's communicative capabilities and enable greater degrees of control and interaction. Furthermore, HD-sEMG is also helpful in elucidating processes contributing to muscle fatigue by providing access to study physiological relevant variables (e.g. propagation velocity) at the global muscle and single motor unit level. The insights gained can be used to devise operational and algorithmic protocols that compensate for fatigue-induced performance variation [77].

HD-sEMG has become a powerful myoelectric imaging tool. The capability to produce detailed view of skeletal muscle activities has enabled high gesture recognition accuracy with large numbers of gestures, with the added benefit of being able to detect muscle fatigue and user weariness. Unfortunately, the hardware complexity involved with such a large number of channels has limited HD-sEMG to benchtop systems in academic research laboratories. An emerging thrust has thus been to miniaturize this technology for more portable sensing applications.

## 2.3 Compact sEMG Systems

Ultimately, the success of sEMG as a ubiquitous NUI is limited by the simplicity, compactness, and comfortability of its physical form. To that end, there has been a significant amount of work reducing the size and complexity of these systems to facilitate end-user applications. An elastic armband is the most common form factor for its ease of on-and-off boarding, greater sensing coverage with circumferential measurement, and firm placement during hand motions. In addition to snug fit, a stand-alone, portable sEMG armband requires four main functional blocks: signal conditioning, signal processing, power supply, and data transmission. To date, there have been numerous research prototypes and commercialized versions of this type of device, such as those shown in Figure 5 [35]–[40].

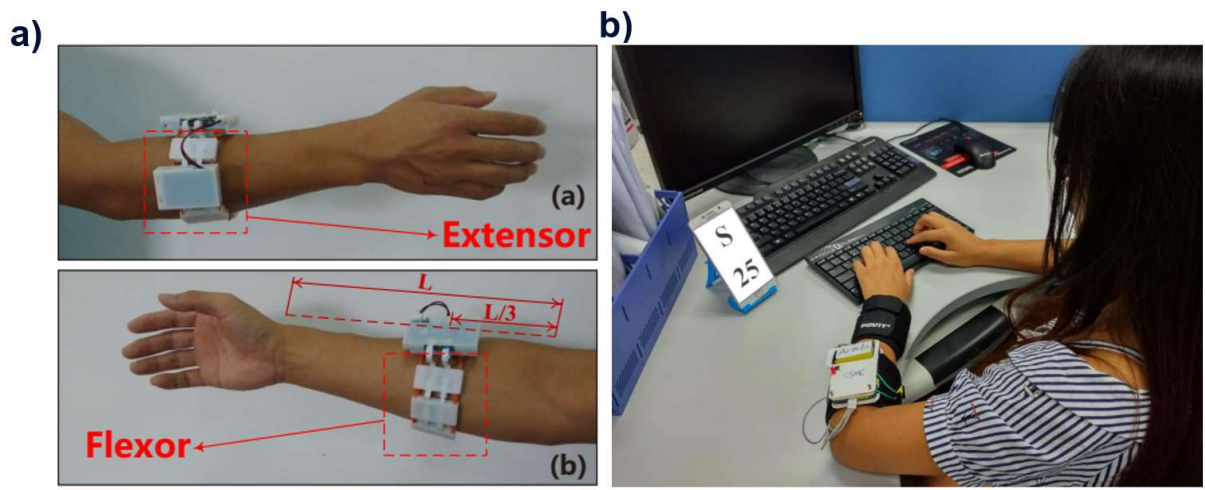

Figure 5: a) Forearm EMG armband capable of wireless gesture detection via onboard electronics. Figures reprinted with permission from [35]. b) ArmIn armband employing sEMG sensor for keystroke detection. Reprinted with permission from [38].

The Myo Armband made by Thalmic Labs (Figure 6a) was the most advanced consumer-grade sEMG device on the market and its low-cost made it a common platform for sEMG application development [78]–[82]. Figure 6b-c presents a tear-down view of the internal layout and electronic circuitry of the Myo Armband. The band consists of a ring of dry metal electrodes held together by elastic plastic, allowing full circumferential electrode placement around the arm and the ability to expand and contract to fit different sized users. Packed inside this device is a low noise operational amplifier circuit directly wired to each of the eight sensing channels, an inertial measurement unit for spatial positional sensing, a Bluetooth module for signal transmission, an ARM microcontroller unit for information processing, and two 3.7-volt Lithium ion batteries providing continuous operation up to 24 hours [34].

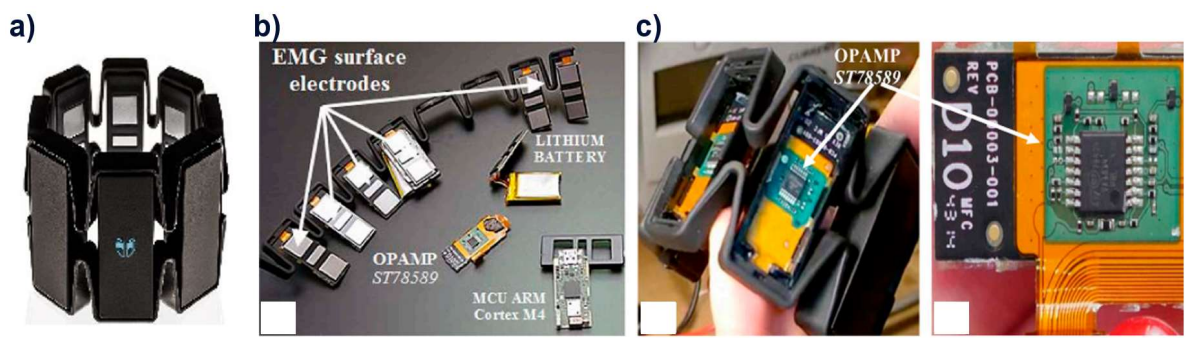

Figure 6: a) Myo Armband by Thalmic Labs, (b) its internal hardware layout and constituents, and (c) blow-up view of one of the sensor modules displaying the amplifier circuit. Figures reprinted from [34] (CC BY-NC-ND 4.0).

Another common form factor employs a distributed sensing strategy where each sEMG sensor acts as a stand-alone "satellite" unit responsible for a specific region of the measurement. Each unit wirelessly transmits its data to a central hub for collection and analysis [83]. An example of this is the FreeMG sensors by BTS Bioengineering [84]; each of these sensors consists of two Ag/AgCl adhesive electrodes for differential sensing and a processing unit measuring 41.5×24.8x14 mm. Each satellite unit weighs 13 grams with a wireless sensing range up to 20 meters and supports continuous operation of 20 units simultaneously up to 6 hours. A distributed sensing strategy offers flexibility in measurement location anywhere on the body without tethering, and they have been shown to enable various gestural detections [85], [86]. While this form is ideal for monitoring major coarse muscle activities, each satellite unit provides only one channel of sEMG measurement, thus high density, detailed measurements on the musculature cannot be collected.

As sEMG methods improve to offer increasingly promising HMI capabilities, portability and wearability become key criteria to enable a comfortable and unobtrusive NUI. Traditional printed circuit board (PCB) integration approaches are severely limiting in this regard, offering a mechanically rigid platform that is at odds with the soft organic tissues of the body. Textile-based systems [42] and printed electronics [87] are two emerging integration techniques that offer a potential solution.

## 2.4 Textile-based sEMG Systems

Despite the development of compact electronics, placement on the body can still be intrusive, inflexible, and uncomfortable because proper skin-contact requires either the application of electrolytic gels in the case of wet electrodes or tight compression for dry metal electrodes [43]. This limitation has motivated the development of smart textiles for increased comfort, reusability, and free body motion [44]. These are fabrics and garments engineered through knitting, weaving, embroidery, and coating techniques to support contact-based sensing applications such as that of sEMG. To date, over 41 articles have been published focusing on electrode material, fabrication methods, and construction approaches that could bring forth long-term, continuous recording of sEMG through textiles. A systemic review of these articles can be found in the report by Guo et al [44]. As an outstanding example, Lee et al demonstrated a multichannel sEMG knit band easily worn by amputees for myoelectric prosthesis [45]. This elastic armband was fabricated by knitting silver-plated conductive yarn as electrodes on a two-layer fabric where the outer layer is non-conductive and moisture-wicking to shield the electrodes in the interior layer from the ambient environment. Shown in Figure 7, snap connectors were sewed into each electrode for sEMG signal transmission. In the form of a stretchable armband, the group showed that despite having a slightly higher averaged electrode-skin impedance (376 ohms) than that of the disposable wet electrode (201 ohms), the knitted electrode armband achieved similar power spectrum characteristics and higher average signal-to-noise ratio (18 dB) than the disposable one (13 dB) according to authors' measurement analysis. To demonstrate its capability in hand gesture detection and recognition, Lee et al. trained a neural network to identify six hand motions and achieved an accuracy up to 93.2% [45]. Although sEMG sensor electronics were not integrated into the knitted armband, this work nonetheless provided the means to obtain high quality skin-electrode connections that is clearly a step towards building a clothing based sEMG system.

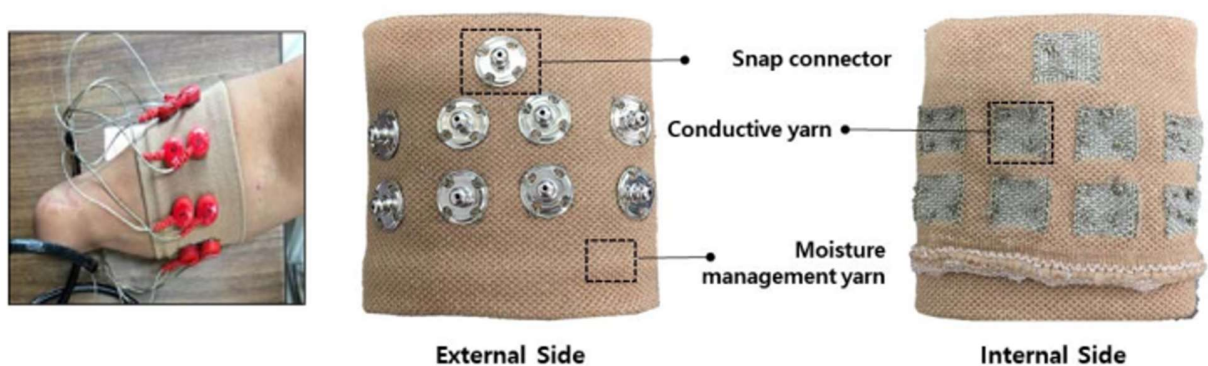

Figure 7: Conductive-thread-based electrode contacts on fabric for EMG data acquisition. Images reprinted with permission from [45].

Besides knitted electrodes, many other promising textile-based flexible sEMG electrodes have been implemented such as by sewing polypyrrole-coated nonwoven fabric sheet as electrodes (PPy-electrode) [46]; dip-coating conductive fabric with highly conductive, hydrophilic Poly(3,4-ethylenedioxythiophene)-poly(styrenesulfonate) (PEDOT:PSS) ink [47]; adapting the use of non-conductive filling pads with conductive fabric to shorten the contact distance between electrode and the skin [48]; and by integrating circular silver electrodes into a stretchable fabric sleeve to detect large area of neuromuscular activities on the forearm [49].

Signal quality obtained from textile-based electrodes is determined by the electrode-skin interfacial impedances. Materials with high electrical conductivity and moisture retention are required to achieve long term stable electrode-skin contact. Unfortunately, there are few conductive fibers or yarns developed specifically for sensing application and commercially available metal-coated yarns are mostly used for anti-static purposes. Further advancement in this space would require the development of dedicated fibers or yarns that are conductive, elastic, and provide moisture control without affecting the wearer's skin hydration state [44]. While data acquisition, processing and transmission electronics are seldom discussed in textile-based sEMG sensors, bulky and tethered systems are typically employed out of convenience and wide availability. It remains to be an area of further development to realize a fully integrated textile sensor where miniaturized electronics can scale and connect seamlessly with textile traces and electrodes. Together with the ongoing development of small footprint sEMG electronics, ordinary garments may one day be augmented with sensory functionality, having little to no wearer awareness of their existence.

## 2.5 Tattoo-based sEMG Systems

Beyond textile integration, there has been a recent rise in research developing tattoo-like epidermal sensors as an unobtrusive form of wearable sEMG. Such a pursuit offers minimal technology footprint and intimate coupling with the skin for long term, high fidelity sensing [50]. Surface electrodes of sEMG systems can take up large sensing area on the body, and the requirement for a gel or water-based interface further presents challenge to miniaturization. Even with dry electrodes, insufficient contact and skin impedances across different individuals under different conditions can result in varying degrees of noise levels. Tattoo-based electrodes have been shown as a promising candidate to overcome these challenges. As the electrode size decreases, greater degree of contact can be achieved with the skin without exerting external pressure through Van der Waals forces to secure them in place. Furthermore, a smaller footprint allows more sensing electrodes to be included for highly refined measurement of biopotentials from minor muscle fibers. One of the earliest works in this area was performed by Lapatki et al. in 2004, in which they constructed a thin, 2D multielectrode grid on a highly flexible polyimide material. After adding silver-chloride coated copper as the electrodes on the grid, the resultant patch is only 470-micron thick and sits on a double-sided adhesive that allows easy attachment to the skin. sEMG measurement was conducted by the group to show differentiable muscle activities on the face [51].

Since this work, tattoo-based systems have advanced significantly. Researchers Kim et al. [52] and Zhou et al. [53] pursued the development of micron thickness, stretchable patches that conform with the human skin and possess electrical performance equivalent to that of gelled electrodes. Using a non-conductive and stretchable laminate (e.g. polydimethylsiloxane) as the base material, conducting metal such as copper or gold are either transferred or sputtered onto the laminate to serve as the electrodes and wiring traces. Additional laminate and etching process finalize the sandwich construct to expose only the electrodes to the skin (Figure 8a). Both research groups have demonstrated excellent sEMG signal quality and the ability to discriminate hand gestures with high accuracy (Figure 8b) [52], [53]. However, there are still a number of challenges with tattoo-like electrodes. For one, the inherent flexibility and soft mechanical properties of these electrodes severely complicate electronics integration. As a result, all demonstrations of epidermal electrodes thus far have used tethered electronic cables to relay data to external processing hardware, limiting the practical usability of these systems. In order to fully realize this type of on-skin sensing form factor, both epidermal contacts via electrodes and the corresponding electronics must scale equally. To this end, an intermediate solution was recently reported by Moin et al.

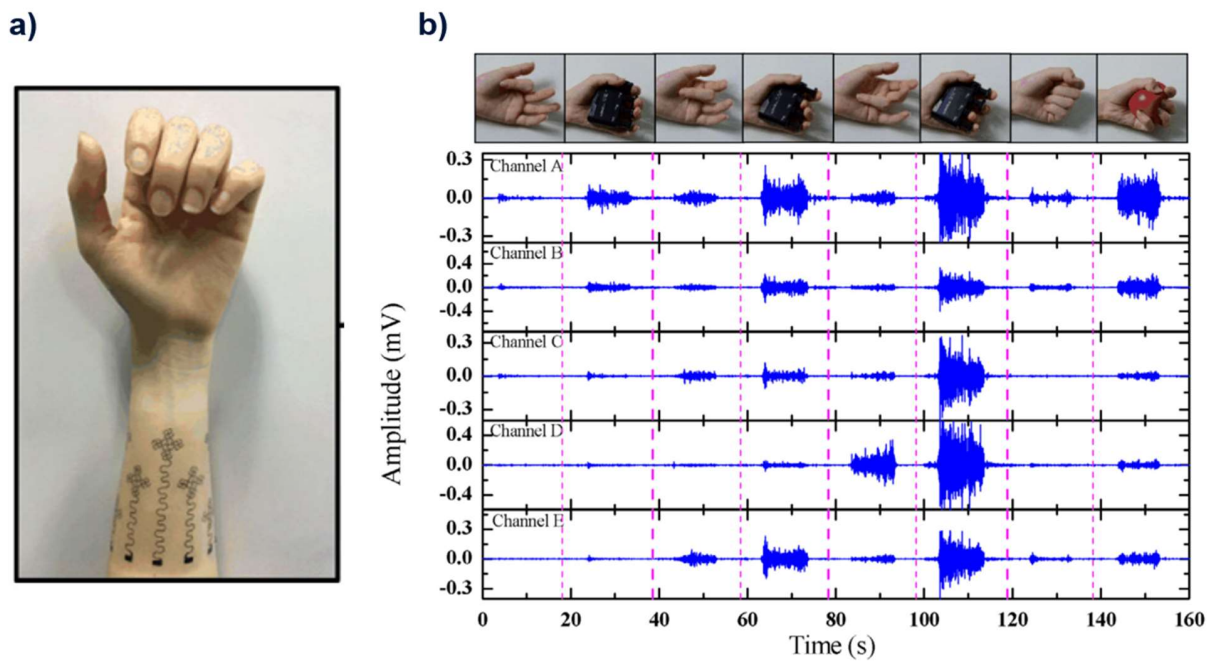

Figure 8: a) Stretchable micron thick sEMG patches attached onto the forearm to record myoelectric activities as shown in b) during various grasping force exertions. Channel A – E are 5 separate EMG recording channels showing EMG signals with dashed pink line separating intervals. Figures reprinted with permission from [53].

[88] which showed electrode arrays screen-printed on a flexible PET substrate to establish a closer, more conformal contact with the skin. With the help of a flexible wire-to-rigid board adaptor, the printed electrode traces were connected with an adjoining, miniaturized electronic board that processed sEMG data and wirelessly transmitted classification results. Unfortunately, this form-factor sacrificed the van der Waals adhesion characteristic of true "tattoo-like" implementations and relied on the application of conductive gel for good skin contact. Nonetheless, the wearable all-in-one arm-wrap with non-negligible interfacial impedance is still an important step towards realizing sEMG as an unobtrusive sensing modality for natural human-machine interaction.

## 2.6 Inertial Measurement Units

sEMG studies across almost all form factors often incorporate inertial measurement units (IMUs), which are electronic devices that measure acceleration and orientation, to improve gesture recognition accuracy. They were first used in combination with sEMG for work in automated sign language recognition. They showed increased accuracy over sEMG-only techniques in Chen et al. 2007 [89] and Kim et al. 2008 [90]. Zhang et al. 2011 [91] was an early work examining fusions of accelerometer and sEMG data, finding increased accuracy for both gesture recognition and Chinese Sign Language recognition.

Accelerometers were then used in Fougner et al. 2011 [92], [93] to compensate for sEMG differences caused by limb angles, after earlier work had found that sEMG activity could vary based on both electrode placement and on joint angles (including joints which are not primarily activated by the muscles being measured). They showed minor improvements in accuracy from the accelerometer, but primarily showed that for a classifier to work at a variety of arm positions, it is necessary to train the classifier in a variety of arm positions.

Later work has continued to show that accelerometer data provides a moderate increase in classification accuracy [94]–[96], typically 2-5%, and that accelerometer-based features contain information distinct from and complementary to the information in forearm-based sEMG [90]. Since they are easily integrated with sEMG sensors, they have become a standard additional sensor in sEMG investigations.

## 3   Information Content of a Natural User Interface

Regardless of the physical implementation of the sEMG system, a natural user interface will require a high-bandwidth physical-to-digital transduction that captures the relevant information needed to accurately infer intent. Before reviewing the digital processing techniques commonly used to do this, we first look at the information content in gestures themselves. We will show that existing user interfaces fall short of what is needed to fully capture the information in natural human gestures and enable a true NUI.

The information bandwidth of human gestures has been studied across many disparate areas, with results typically reported in terms of "actions per second" or "bits per second". Actions per second measures the rate of discrete actions without concern for the information content of each action. For highly trained users, this rate is consistently reported in the range of 5-10 actions per second. For example, in typing, a 168,000-volunteer online study [96] found a 90th-percentile typing speed of 78 words per minute, corresponding to about 6.5 characters per second. The speed of professional gamers such as StarCraft 2 players often average approximately 5 user-interface actions per second, with spikes up to about 8 per second [97]. Native speakers of American Sign Language produce 2-3 signs per second [98], [99] with signs typically comprising at least three distinct "phoneme" elements [100], [101].

In the field of user interface design, Fitts' Law [102] is commonly used to study the information content of interfaces, particularly in the context of pointing and selection tasks [103]. Evaluations of existing gesture interfaces show that users achieve an average of 1.5-2 bit/s with gesture modalities, compared to 4-8 bit/s for mouse input and 6-14 bit/s for direct touchscreen inputs [104], [105]. In contrast, the information content of the motions themselves has been estimated at 10 bit/s for arm movement, 23 bit/s for wrist movement, and 38 bit/s for finger movement [106]. Existing interfaces are not even close to matching the rate of whole-arm systems, much less taking advantage of the higher information bandwidth of individual finger movements.

The ability to accurately capture information from gestures partially depends on the sampling frequency of the data acquisition hardware. In general, the higher the sampling rate, the greater the digital reconstitution of the physical signal. As discussed in section 1.4, the power spectrum of EMG signals is below 500 Hz. The Nyquist theorem therefore dictates that a sampling rate of 1 kHz or higher is necessary to capture the muscle signal faithfully. This is commonly observed with lab-based sEMG experiments where the majority of experimental sEMG systems employ a sampling frequency of 1 kHz or higher. As a special instance, the commercial availability of the low-cost Myo Armband (see §2.3), sampling at 200 Hz, have led to success of some gestural detection using ingenious techniques, but fell significantly short when it comes to detection of large number of gestures and recognition accuracy [78]–[82], [107]. Researchers typically report that accuracy begins degrading when sampling rates fall below a threshold; the exact threshold has varied between reports, including 200 Hz [108], 400 Hz [109], and 600 Hz [110].

In sEMG, the number of data acquisition channels has a large impact on the amount of information captured, with higher channel count systems consistently yielding a larger number and greater subtlety of gestures recognized. Selecting electrode signals on the basis of mutual information, Amma et al. [76] found continuing improvement in recognition accuracy until approximately 100 electrodes per arm. Conversely, Li et al. [110] found improvement up to 20 channels in a study of post-stroke subjects.

A review of the number of gestures classified using sEMG versus number of channels employed from 88 different publications is shown in Figure 9a. For work that reported classification across multiple time windows, accuracies for time windows between 150 ms – 200 ms were recorded as this was the most universal window length. In general, work that used larger numbers of channels were able to classify a larger number of gestures with greater accuracy. Classifying Sign Language gestures is a notable exception, with as many as 72 Signs being correctly classified at 96% accuracy using only 5 sEMG channels [91], which may be due to a heavy reliance on IMU's in Sign Language classification tasks. Excluding Sign Language work and reports with classification accuracy below 90% results in a clear logarithmic trend of identifiable gestures versus channel count (Figure 9b). This suggests that while higher number of channels is desirable, there are diminishing returns for increasingly dense physical implementations. In addition, while the overall trend is clear, signal processing technique plays a key role in classification accuracy on any single dataset. In the next section, we review traditional digital signal processing techniques used to process and classify sEMG signals as well as more recent work using deep learning techniques, such as convolutional neural network (CNNs).

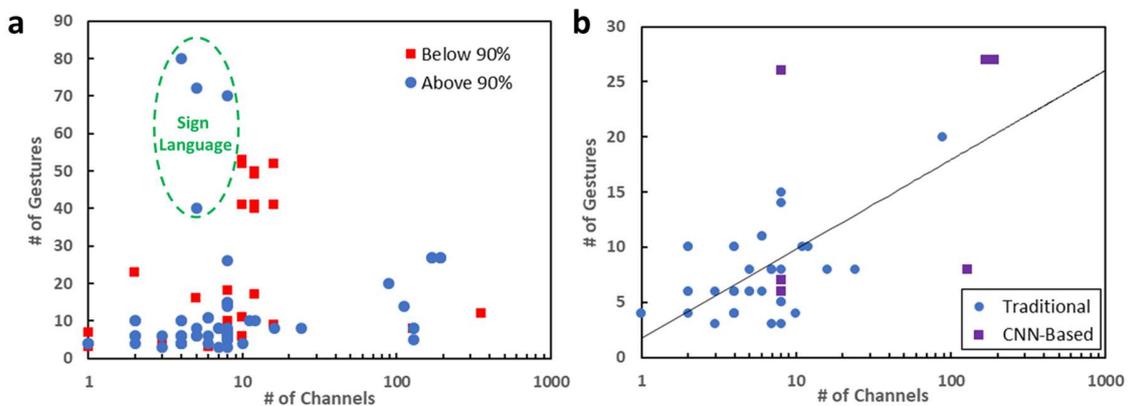

Figure 9: a) A review of 88 articles summarizing the number of gestures detected and the number of sEMG channels employed. Work that reported classification accuracy greater than 90% shown in blue, less than 90% in red. b) Subset of sEMG work where greater than 90% classification accuracy is reported, excluding work on sign language. Reports using CNN-based classification are highlighted in purple. Solid black line depicts logarithmic trendline.

# 4   sEMG Signal Processing

The first stage of interpreting sEMG signals is preprocessing the raw digitized waveforms by normalizing the data and removing powerline noise and DC offset. We will not cover preprocessing techniques in detail, instead directing the reader to Hudgins, Parker, and Scott 1993 [111] and to Englehart, Hudgins, and Parker 2001 [112], which presented preprocessing methods that became standard in the literature. Rodríguez-Tapia et al. [18] provides a meta-analysis of recent signal collection and preprocessing methods in deep learning-based sEMG applications, covering articles published from 2014 to 2018.

Most research systems focus on recognizing individual gestures one at a time [113]–[115]. After preprocessing, the traditional pipeline for gesture recognition via sEMG extracts a number of pre-chosen *"features"* which characterize the gesture in a much more compact way than the original signal, while still retaining the information necessary for gesture recognition. In parallel, a *gesture detection* method is typically used to determine when the continuous signal contains a gesture. When a gesture is recognized, a *classification* method is used to determine which of the set of gestures it is. A typical system architecture is presented in Figure 10.

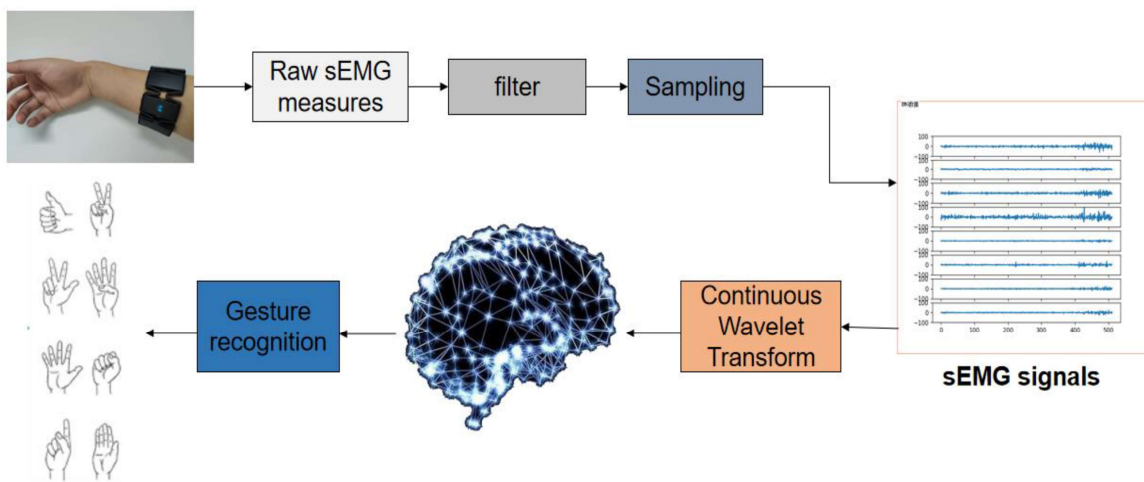

Figure 10: One typical gesture recognition system architecture, from [80]. sEMG signals are collected, the frequency range of interest is selected via filtering or other preprocessing, and the signals are then digitally sampled. *Features* of interest (in this case Continuous Wavelet Transform coefficients) are calculated from the signals, and then a *classifier* (in this case a neural network) is applied to these features, outputting the probabilities of various gestures. Figure reproduced from [80] (CC BY 4.0).

It is important to note that sEMG signals vary widely from person to person, and can significantly differ even between the same user on different days [116]. When a system is evaluated on a user who was not part of the training data, this is the "interuser" accuracy. When a system is evaluated on a known user who was part of the training data, but the device has been removed and re-placed on the user, this is the "intersession" accuracy. When some of the gesture repetitions from a user and session are used for training a system, and the system is evaluated on other repetitions from this user and session, this is the "intrasession" accuracy. The intrasession accuracy, intersession accuracy, and interuser accuracy form a hierarchy of increasing difficulty. When comparing accuracy numbers from different authors on the same dataset, it is important to keep in mind which accuracy is being reported. One approach to achieving higher interuser accuracy is to "pre-train" the system on a large training set from multiple users, then to fine-tune the system on a limited training set from an individual user [117], [118]. Other authors have focused on features and learning techniques specialized for generalizing between sessions or between users [119].

## 4.1  Feature Selection

Traditional machine learning techniques require excessive amounts of training data and computation time as input data becomes larger. They typically operate well on inputs with several dozen features, rather than on input vectors of tens of thousands, as are generated by EMG sensors. Systems using pre-deep-learning techniques thus typically rely on separating the input signal into time windows [120], [121], and cleverly designing a small number of *features* which characterize the signal in each window. This also addresses the mismatch between the natural input rate of an EMG gesture recognition systems (1 kHz-4 kHz, with 32 or more channels) and the natural rate of gestures (average 5 gestures per second, or 10-20 bits per second).

For example, for each channel, features might include the mean of the absolute value of that channel's readings within a time window, or the number of times the signal crosses zero within that window. Early work in EMG gesture recognition tended to focus more on these feature selection aspects than on the classifiers themselves. A large number of reports have stated that the choice of features was far more important to classification accuracy than the algorithms used afterwards to classify these features [111], [114], [120], [122], [123]. Some of the modern deep learning techniques discussed in later sections do operate directly on large dimensionality sEMG data.

Several reviews of sEMG features exist, and they have been extensively documented by Zardoshti-Kermani et al. [124], Boostani and Moradi 2003 [125], and Phinyomark, Phukpattaranont, and Limsakul 2012 [121].  Hudgins, Parker, and Scott 1993 [111] introduced a set of time-domain features which became standard; later, Englehart, Hudgins, and Parker 2001 [112] provided a particularly influential set of wavelet-based features, which were extended by many works that have focused on the suitability of features for more specific topics, including intersession recognition [122], [126], intersubject recognition [127], and low-frequency sensors [128].

After features have been created, dimensionality reduction techniques are often used to turn a medium-sized set of features into a smaller set that is expected to be nearly as good [120]. Dimensionality reduction in machine learning is a large topic; in the sEMG space, researchers have used general techniques such as principal component analysis (PCA) [112], [123] and linear discriminate analysis (LDA) [122], as well as more specific techniques for sEMG reported by Khushaba and Kodagoda [129]. These techniques have largely been superseded by later deep learning techniques, which are discussed later.

## 4.2  Gesture Detection

In practical sEMG applications, much of the user's time will be spent *between* hand or finger gestures; building a NUI requires detecting when and whether gestures have happened, so that the system can react appropriately. Since gesture detection is often done separately from gesture classification, some research focuses solely on the issue of classification, assuming that gesture recognition has been done. For example, a paper might study the question of how to classify a 300 ms sEMG signal which is guaranteed to contain exactly one gesture. While this work cannot be applied directly to streaming sEMG data, it is still helpful when used as part of a larger system. Additionally, many sets of training data (e.g. NinaPro) are in exactly such a form; they were collected by asking volunteers to do one gesture at a time, at known times (see §4.5).

If data is not *a priori* arranged in single-gesture windows, one approach is to explicitly perform gesture detection, either as a preprocessing step or simply in parallel with classification. sEMG amplitudes track muscle contraction; gestures will thus correspond to continuous periods of high-amplitude signals. We can look at the power envelope and classify the start of a gesture when the power goes over a certain threshold. This approach appeared early in the literature [111], and is used by several teams [75], [76], particularly in analyzing the CSL-hdemg dataset (see §4.5), where recorded gestures occur sometime within a 3-second window.

Another alternative is to train the "rest" or "null" gesture directly, as one of the gestures to recognize [91], [109]. This is a natural approach, but care must be taken when analyzing the performance of such a system, to ensure that accuracy statistics are balanced between different classes. For example, in the first NinaPro dataset, over 50% of all samples are "rest" samples, corresponding to the time between gestures [130]. An algorithm that could successfully differentiate between "rest" and "gesture" could thus appear to yield a minimum of 50% "accuracy", even without differentiating between different types of gestures.

For high-rate gestures, such as the maximum rate of 10 gestures a second discussed in Section 3, each gesture would occupy < 100 ms of time. However, most gesture detection and classification algorithms work by allowing for a single classification within a predefined window of time, typically between 150 ms – 200 ms, that is too long to capture such a high rate. Conversely, for applications such as industrial control and slow-moving prosthetics, it may be desirable to avoid false positives, and the device being controlled may well have a slower response rate than the speed of gesture recognition. In this setting, postprocessing will often be appropriate to avoid responding to transient signals or responding in error. Methods such as taking a majority vote of successive time windows may thus be appropriate and yield significantly higher accuracies [112], [115], [131].

## 4.3 Classifiers

After feature extraction, an input of tens of thousands of samples per second has been reduced to a few dozen. The training data can then be used to train a *classifier*, to distinguish between the gestures in question. Traditional machine learning techniques used at this point have included Linear Discriminant Analysis (LDA) [92], [108], [115], [123], Support Vector Machines (SVM) [90], [95], [108], [116], [127], neural network models [111], [116], [123], decision trees [91], [95], hidden Markov models [91], K Nearest Neighbor [90], [95], [108], Gaussian mixture models [132], Kernel Regularized Least Squares [94], naive Bayes classifier [76], [95] and Locally Weighted Projection Regression (LWPR) [116]. LDA was the most common general-purpose classifier used for gesture recognition [115], but more recent work has focused on automatically constructing classifiers through deep learning. As shown in Figure 9.b, deep learning based systems, particularly CNNs, have outperformed other classifiers in terms of the number of gestures that can be accurately identified for a given number of channels.

## 4.4 Signal Interpretation Through Deep Learning

"Deep learning" refers to the use of multi-layered neural networks to simultaneously perform both feature extraction and classification. The ability to perform deep learning on practical datasets rapidly evolved in the 2010s as a result of both hardware and algorithmic improvements that made it feasible to train many-layered neural networks on increasingly large sets of training data. When deep networks are set up correctly, the "lowest" layers of the input extract features directly from the input signal. Successively higher levels use these features to build more complicated composite features, until the final layers are able to easily classify the signal using the features constructed [133].

The most successful neural network architectures have been those which take advantage of the structure of the input. For example, a 200x200 pixel grayscale image could be viewed as simply a list of 40,000 independent numbers, but that view would discard important information about the geometric layout of the image; neural networks which take that view are very difficult to train and use far too many parameters. The fact that the pixels occur in a two-dimensional layout is fundamental to building high-performance neural networks for image analysis [134], [135].

In the case of EMG signals, there are several forms of structure available. EMG signals contain one time dimension, and one or two spatial dimensions (depending whether the electrodes are arranged linearly or in a grid). In the spatial dimension(s), networks commonly use *convolutional* layers, which learn features that are applied in a sliding-window fashion across the one or two spatial dimensions [75], [117], [118], [131]. In the time dimension, researchers have worked with *recurrent* layers, which learn features that read an input sequentially, "remembering" some of the values it has seen. The most popular type of recurrent layer is the "Long Short-Term Memory" (LSTM) layer, which explicitly models the process of learning which values to "forget" [136]. Researchers have also used convolutional layers to analyze time series data.

Due to the physical mechanisms that generate sEMG signals (see Section 1.2), they are highly stochastic and historically it was generally accepted that the instantaneous value of an EMG signal was of little use [111], [137]. However, on HD-sEMG signals, which have two spatial dimensions, the spatial patterns in an instantaneous reading have been analyzed successfully with high accuracy via convolutional networks [75], [114], using techniques from image recognition (see Figure 11). A similar architecture has also been used for analyzing sparse sEMG signals through an entire time window, using a 2-dimensional convolution that looks for patterns over both the time and space axes [114], [131], [133], [138]–[140]. One recent paper makes the natural extension to a 3-dimensional convolution over an HD-sEMG signal evolving through time [141].

LSTMs are sometimes used alone for time-series analysis [81], but they are typically combined with convolutions, also in the time dimension. Most research combining the two uses the convolutional layers for feature extraction, with LSTM layers being used to build the classifier [140], [142]–[145], but there is limited evidence that the other order may work equivalently well [107].

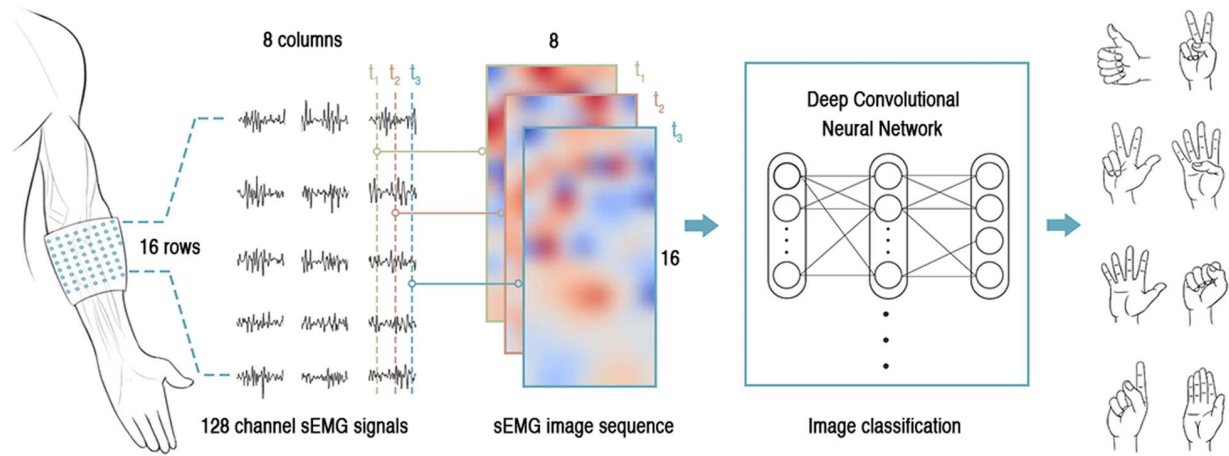

Figure 11, showing the architecture of the deep learning network of [75]. Electrodes are placed in a square grid pattern, and the instantaneous sEMG signal amplitudes are then processed using networks originally designed for image processing applications. Figure reproduced from [75]. (CC BY 4.0)

## 4.5 Public Datasets

Machine learning systems often require large amounts of training data and comparing different learning strategies is much easier when they are applied to the same dataset. Publicly available datasets are widely used in the sEMG space. Some of the available gesture recognition datasets were originally collected for prosthetics applications, but for ethical and practical reasons many of these are nonetheless collected on intact subjects, which are often a reasonable proxy for amputees [146].

Early datasets focused on collecting large numbers of gestures for sparse electrodes, and include an unnamed collection by Khushaba et al. [60], [129], [137], [147]–[149] and datasets collected by the NinaPro Project [150]–[152]. The Khushaba datasets include sets focusing on static gesture recognition and on recognizing level of grasp force and are sampled at high frequencies (2–4kHz). The NinaPro datasets focus on general sEMG gesture recognition, are larger (in number of gestures and subjects) than most, and are widely used in the literature; they often include accelerometer data. The putEMG dataset [61] instead focuses on a smaller number of gestures (8), but with a large number of participants (44) and repetitions (40 over 2 sessions).

In the HD-sEMG space, the CSL-hdemg dataset [76] and the CapgMyo dataset [153] are widely used. CSL-hdemg contains 192 electrodes measuring 168 channels at 2048Hz, measuring 27 gestures on 5 subjects, and is suitable for general hd-sEMG gesture recognition. Each subject is recorded for 5 sessions, and each session contains 10 repetitions of each gesture. CapgMyo uses 128-channel sEMG data at 1000Hz on 23 subjects, and records a subset of the NinaPro gestures, focusing on intersession and intersubject recognition.

## 5 Conclusion & Future Outlook

The development of sEMG over the last few decades has advanced it towards a natural wearable form factor, which emphasizes user comfort and provides unrestricted bodily movement. While striving to achieve a natural user experience, there is an implicit balance between the simplicity of a low-channel count system and an information-rich but more complicated HD-sEMG. The incorporation of high-dimensional arrays of electrodes through epidermal tattoo-based contact methods and textile-based systems has slowly increased the number of channels deployable while maintaining user comfort. Textile embedded sensors are perhaps the most convenient and practical implementation but are still relatively limited until further reduction of electrode footprint on fabric can be realized. So far, only up to 8-channels have been reported in textile-based sEMG systems [45] and the ultimate limit of textile electrode integration density remains to be investigated. Tattoo-based systems, on the other hand, are still limited by the delicate integration of soft-materials with traditional rigid electronics to complete a fully stand-alone system.

Regardless of the work still needed to be done on the physical implementation, existing hardware systems have allowed for a rapid advance in signal processing techniques. Traditional EMG processing techniques rooted from medical diagnostics has slowly evolved and has recently embraced deep learning approaches such as CNNs and LSTMs. These new algorithms have demonstrated record-high number of accurately classified gestures, while also reducing latency and processing time from traditional techniques. Some systems have demonstrated high accuracy (>99%) classification of gestures with just 40 ms of acquisition [140], however they rely on exceedingly large numbers of acquisition channels. To enable a truly fluid and natural user interface, future work must either further increase accuracy at low latencies (< 100 ms) using low channel count hardware or wait for hardware integration approaches to provide the larger channel counts required of current algorithms.

Finally, as sEMG hardware and classification systems continue to provide higher accuracy at lower latencies, the overarching systems that map the outcome of these gestures to actions still needs to be fully investigated. To date, most systems use individual gestures with one-to-one control mappings, such as up/down/left/right to control a drone and are not truly a natural way to interact with a machine. Natural gesturing can take many forms and can have a diverse number of meanings depending on situational context. Incorporating context into an sEMG system and developing a system that can understand the language of natural human gesture is still a very open and exciting research area, and may be the ultimate hurdle to overcome before sEMG systems can be ubiquitously deployed as natural human-machine interfaces.

# 6 References

[1]     P. Papcun, E. Kajáti, and J. Koziorek, "Human machine interface in concept of industry 4.0," in *DISA 2018 - IEEE World Symposium on Digital Intelligence for Systems and Machines, Proceedings*, Aug. 2018, pp. 289–296, doi: 10.1109/DISA.2018.8490603.

[2]     G. Johannsen, "Human-Machine Systems Research for Needs in Industry and Society," *IFAC Proc. Vol.*, vol. 34, no. 16, pp. 1–9, Sep. 2001, doi: 10.1016/S1474-6670(17)41493-5.

[3]     K. Wucherer, "HMI, The Window to the Manufacturing and Process Industry," *IFAC Proc. Vol.*, vol. 34, no. 16, pp. 101–108, Sep. 2001, doi: 10.1016/S1474-6670(17)41508-4.

[4]     V. Villani, L. Sabattini, J. N. Czerniaki, A. Mertens, B. Vogel-Heuser, and C. Fantuzzi, "Towards modern inclusive factories: A methodology for the development of smart adaptive human-machine interfaces," in

*2017 22nd IEEE International Conference on Emerging Technologies and Factory Automation (ETFA)*, Sep. 2017, pp. 1–7, doi: 10.1109/ETFA.2017.8247634.

[5] M. K. Weldon, *The future X network: a Bell Labs perspective*. CRC press, 2016.

[6] H. Viswanathan and P. E. Mogensen, "Communications in the 6G Era," *IEEE Access*, vol. 8, pp. 57063–57074, 2020, doi: 10.1109/ACCESS.2020.2981745.

[7] E. Palermo, J. Laut, O. Nov, P. Cappa, and M. Porfiri, "A natural user interface to integrate citizen science and physical exercise," *PLoS One*, vol. 12, no. 2, p. e0172587, Feb. 2017, doi: 10.1371/journal.pone.0172587.

[8] N. Pavón-Pulido, J. A. López-Riquelme, and J. J. Feliú-Batlle, "IoT Architecture for Smart Control of an Exoskeleton Robot in Rehabilitation by Using a Natural User Interface Based on Gestures," *J. Med. Syst.*, vol. 44, no. 9, p. 144, Sep. 2020, doi: 10.1007/s10916-020-01602-w.

[9] R. A. Suarez Fernandez, J. L. Sanchez-Lopez, C. Sampedro, H. Bavle, M. Molina, and P. Campoy, "Natural user interfaces for human-drone multi-modal interaction," in *2016 International Conference on Unmanned Aircraft Systems (ICUAS)*, Jun. 2016, pp. 1013–1022, doi: 10.1109/ICUAS.2016.7502665.

[10] J. Grubb and J. Cohn, "The Evolution of Human Systems: A Brief Overview," Springer, Berlin, Heidelberg, 2011, pp. 60–66.

[11] N. O-larnnithipong, A. Barreto, S. Tangnimitchok, and N. Ratchatanantakit, "Orientation correction for a 3D hand motion tracking interface using inertial measurement units," in *Lecture Notes in Computer Science (including subseries Lecture Notes in Artificial Intelligence and Lecture Notes in Bioinformatics)*, Jul. 2018, vol. 10903 LNCS, pp. 321–333, doi: 10.1007/978-3-319-91250-9_25.

[12] B. Schabron, A. Reust, J. Desai, and Y. Yihun, "Integration of Forearm sEMG Signals with IMU Sensors for Trajectory Planning and Control of Assistive Robotic Arm," in *Proceedings of the Annual International Conference of the IEEE Engineering in Medicine and Biology Society, EMBS*, Jul. 2019, pp. 5274–5277, doi: 10.1109/EMBC.2019.8856699.

[13] C. J. Wu, K. H. Lin, M. L. Hsieh, and J.-Y. Y. Chang, "Realization of natural user interface for computer control with KNN classifier enhanced smart glove," Jun. 2019, doi: 10.1115/ISPS2019-7493.

[14] L. Almeida *et al.*, "Towards natural interaction in immersive reality with a cyber-glove," in *Conference Proceedings - IEEE International Conference on Systems, Man and Cybernetics*, Oct. 2019, vol. 2019-Octob, pp. 2653–2658, doi: 10.1109/SMC.2019.8914239.

[15] S. Kean, J. C. Hall, and P. Perry, "Behind the Technology," in *Meet the Kinect*, Berkeley, CA: Apress, 2011, pp. 29–44.

[16] S. Kean, J. C. Hall, and P. Perry, "Application Development with the Beckon Framework," in *Meet the Kinect*, Berkeley, CA: Apress, 2011, pp. 101–127.

[17] F. Malawski and J. Gałka, "System for multimodal data acquisition for human action recognition," *Multimed. Tools Appl.*, vol. 77, no. 18, pp. 23825–23850, Sep. 2018, doi: 10.1007/s11042-018-5696-z.

[18] B. Rodríguez-Tapia, I. Soto, D. M. Martinez, and N. C. Arballo, "Myoelectric Interfaces and Related Applications: Current State of EMG Signal Processing–A Systematic Review," *IEEE Access*, vol. 8, pp. 7792–7805, 2020, doi: 10.1109/ACCESS.2019.2963881.

[19] S. Lee *et al.*, "Wireless Epidermal Electromyogram Sensing System," *Electronics*, vol. 9, no. 2, p. 269, Feb. 2020, doi: 10.3390/electronics9020269.

[20] A. Phinyomark and E. Scheme, "EMG Pattern Recognition in the Era of Big Data and Deep Learning," *Big Data Cogn. Comput.*, vol. 2, no. 3, p. 21, Aug. 2018, doi: 10.3390/bdcc2030021.

[21] P. Kumari, L. Mathew, and P. Syal, "Increasing trend of wearables and multimodal interface for human activity monitoring: A review," *Biosensors and Bioelectronics*, vol. 90. Elsevier Ltd, pp. 298–307, Apr. 15, 2017, doi: 10.1016/j.bios.2016.12.001.

[22] T. Ray *et al.*, "Soft, skin-interfaced wearable systems for sports science and analytics," *Curr. Opin. Biomed. Eng.*, vol. 9, pp. 47–56, Mar. 2019, doi: 10.1016/J.COBME.2019.01.003.

[23] T. Moritani, D. F. Stegeman, and R. Merletti, "Basic physiology and biophysics of EMG signal generation," *Electromyogr. Physiol. Eng. Noninvasive Appl.*, pp. 1–20, 2004.

[24] M. Barbero, R. Merletti, and A. Rainoldi, *Atlas of Muscle Innervation Zones*. Milano: Springer Milan, 2012.

[25] D. Lacomis, "Electrodiagnostic approach to the patient with suspected myopathy.," *Neurol. Clin.*, vol. 30, no. 2, pp. 641–60, May 2012, doi: 10.1016/j.ncl.2011.12.007.

[26] V. Agostini, M. Ghislieri, S. Rosati, G. Balestra, and M. Knaflitz, "Surface Electromyography Applied to Gait Analysis: How to Improve Its Impact in Clinics?," *Front. Neurol.*, vol. 11, p. 994, Sep. 2020, doi: 10.3389/fneur.2020.00994.

[27] J.-W. Jeong *et al.*, "Materials and Optimized Designs for Human-Machine Interfaces Via Epidermal Electronics," *Adv. Mater.*, vol. 25, no. 47, pp. 6839–6846, Dec. 2013, doi: 10.1002/adma.201301921.

[28] R. Merletti, A. Botter, and U. Barone, "Detection and Conditioning of Surface EMG Signals," in *Surface Electromyography : Physiology, Engineering, and Applications*, Hoboken, New Jersey: John Wiley & Sons, Inc., 2016, pp. 1–37.

[29] C. J. De Luca, "The Use of Surface Electromyography in Biomechanics," *J. Appl. Biomech.*, vol. 13, no. 2, pp. 135–163, May 1997, doi: 10.1123/jab.13.2.135.

[30] G. L. Soderberg, "Recording Techniques," in *Selected topics in surface electromyography for use in the occupational setting: expert perspectives*, DHHS (NIOS., U.S. Department of Health and Human Services, 1992, pp. 24–43.

[31] T. W. Beck *et al.*, "A comparison of monopolar and bipolar recording techniques for examining the patterns of responses for electromyographic amplitude and mean power frequency versus isometric torque for the vastus lateralis muscle," *J. Neurosci. Methods*, vol. 166, no. 2, pp. 159–167, Nov. 2007, doi: 10.1016/j.jneumeth.2007.07.002.

[32] D. A. Winter, *Biomechanics and Motor Control of Human Movement*. Hoboken, NJ, USA: John Wiley & Sons, Inc., 2009.

[33] M. Rojas-Martínez, M. A. Mañanas, and J. F. Alonso, "High-density surface EMG maps from upper-arm and forearm muscles," *J. Neuroeng. Rehabil.*, vol. 9, no. 1, p. 85, Dec. 2012, doi: 10.1186/1743-0003-9-85.

[34] P. Visconti, F. Gaetani, G. A. Zappatore, and P. Primiceri, "Technical features and functionalities of Myo armband: An overview on related literature and advanced applications of myoelectric armbands mainly focused on arm prostheses," *Int. J. Smart Sens. Intell. Syst.*, vol. 11, no. 1, pp. 1–25, Jan. 2018, doi: 10.21307/ijssis-2018-005.

[35] X. Tang, Y. Liu, C. Lv, and D. Sun, "Hand Motion Classification Using a Multi-Channel Surface Electromyography Sensor," *Sensors*, vol. 12, no. 2, pp. 1130–1147, Jan. 2012, doi: 10.3390/s120201130.

[36] T. S. Saponas, D. S. Tan, D. Morris, J. Turner, and J. A. Landay, "Making muscle-computer interfaces more practical," in *Proceedings of the 28th international conference on Human factors in computing systems - CHI '10*, 2010, p. 851, doi: 10.1145/1753326.1753451.

[37] L. P. Simmons and J. S. Welsh, "Compact human-machine interface using surface electromyography," in *2013 IEEE/ASME International Conference on Advanced Intelligent Mechatronics*, Jul. 2013, pp. 206–211, doi:

10.1109/AIM.2013.6584093.

[38] Q. Yang, Y. Zou, M. Zhao, J. Lin, and K. Wu, "ArmIn: Explore the feasibility of designing a text-entry application using EMG signals," in *ACM International Conference Proceeding Series*, Nov. 2018, pp. 117–126, doi: 10.1145/3286978.3287030.

[39] Weichao Guo, Xinjun Sheng, Jianwei Liu, Lei Hua, Dingguo Zhang, and Xiangyang Zhu, "Towards zero training for myoelectric control based on a wearable wireless sEMG armband," in *2015 IEEE International Conference on Advanced Intelligent Mechatronics (AIM)*, Jul. 2015, pp. 196–201, doi: 10.1109/AIM.2015.7222531.

[40] L. Popović Maneski, I. Topalović, N. Jovičić, S. Dedijer, L. Konstantinović, and D. B. Popović, "Stimulation map for control of functional grasp based on multi-channel EMG recordings," *Med. Eng. Phys.*, vol. 38, no. 11, pp. 1251–1259, Nov. 2016, doi: 10.1016/J.MEDENGPHY.2016.06.004.

[41] Xing Shusong and Zhang Xia, "EMG-driven computer game for post-stroke rehabilitation," in *2010 IEEE Conference on Robotics, Automation and Mechatronics*, Jun. 2010, pp. 32–36, doi: 10.1109/RAMECH.2010.5513218.

[42] Z. Lou, L. Wang, and G. Shen, "Recent Advances in Smart Wearable Sensing Systems," *Advanced Materials Technologies*, vol. 3, no. 12. John Wiley & Sons, Ltd, p. 1800444, Dec. 2018, doi: 10.1002/admt.201800444.

[43] G. Acar, O. Ozturk, A. J. Golparvar, T. A. Elboshra, K. Böhringer, and M. Kaya Yapici, "Wearable and flexible textile electrodes for biopotential signal monitoring: A review," *Electronics*, vol. 8, no. 5, p. 479, Apr. 2019, doi: 10.3390/electronics8050479.

[44] L. Guo, L. Sandsjö, M. Ortiz-Catalan, and M. Skrifvars, "Systematic review of textile-based electrodes for long-term and continuous surface electromyography recording," *Textile Research Journal*, vol. 90, no. 2. SAGE PublicationsSage UK: London, England, pp. 227–244, Jan. 2020, doi: 10.1177/0040517519858768.

[45] S. Lee, M. O. Kim, T. Kang, J. Park, and Y. Choi, "Knit Band Sensor for Myoelectric Control of Surface EMG-Based Prosthetic Hand," *IEEE Sens. J.*, vol. 18, no. 20, pp. 8578–8586, Oct. 2018, doi: 10.1109/JSEN.2018.2865623.

[46] Y. Jiang, S. Sakoda, M. Togane, S. Morishita, B. Lu, and H. Yokoi, "A highly usable and customizable sEMG sensor for prosthetic limb control using polypyrrole-coated nonwoven fabric sheet," in *2015 IEEE SENSORS - Proceedings*, Nov. 2015, pp. 1–4, doi: 10.1109/ICSENS.2015.7370380.

[47] A. Niijima, T. Yamada, T. Isezaki, R. Aoki, and T. Watanabe, "hitoeCap: Wearable EMG sensor for monitoring masticatory muscles with PEDOT-PSS textile electrodes," in *Proceedings - International Symposium on Wearable Computers, ISWC*, 2017, vol. Part F1305, pp. 215–220, doi: 10.1145/3123021.3123068.

[48] B. Sumner, C. Mancuso, and R. Paradiso, "Performances evaluation of textile electrodes for EMG remote measurements," in *Proceedings of the Annual International Conference of the IEEE Engineering in Medicine and Biology Society, EMBS*, Jul. 2013, pp. 6510–6513, doi: 10.1109/EMBC.2013.6611046.

[49] M. Gazzoni, N. Celadon, D. Mastrapasqua, M. Paleari, V. Margaria, and P. Ariano, "Quantifying forearm muscle activity during wrist and finger movements by means of multi-channel electromyography," *PLoS One*, vol. 9, no. 10, p. e109943, Oct. 2014, doi: 10.1371/journal.pone.0109943.

[50] S. Kabiri Ameri *et al.*, "Graphene Electronic Tattoo Sensors," *ACS Nano*, vol. 11, no. 8, pp. 7634–7641, Aug. 2017, doi: 10.1021/acsnano.7b02182.

[51] B. G. Lapatki, J. P. van Dijk, I. E. Jonas, M. J. Zwarts, and D. F. Stegeman, "A thin, flexible multielectrode grid for high-density surface EMG," *J. Appl. Physiol.*, vol. 96, no. 1, pp. 327–336, Jan. 2004, doi: 10.1152/japplphysiol.00521.2003.

[52] N. Kim, T. Lim, K. Song, S. Yang, and J. Lee, "Stretchable Multichannel Electromyography Sensor Array Covering Large Area for Controlling Home Electronics with Distinguishable Signals from Multiple Muscles,"

*ACS Appl. Mater. Interfaces*, vol. 8, no. 32, pp. 21070–21076, Aug. 2016, doi: 10.1021/acsami.6b05025.

[53] Y. Zhou, Y. Wang, R. Liu, L. Xiao, Q. Zhang, and Y. Huang, "Multichannel noninvasive human-machine interface via stretchable μm thick sEMG patches for robot manipulation," *J. Micromechanics Microengineering*, vol. 28, no. 1, p. 014005, Jan. 2018, doi: 10.1088/1361-6439/aa9c2e.

[54] J. Kim, S. Mastnik, and E. André, "EMG-based hand gesture recognition for realtime biosignal interfacing," in *Proceedings of the 13th international conference on Intelligent user interfaces - IUI '08*, 2008, p. 30, doi: 10.1145/1378773.1378778.

[55] H. Li, X. Chen, and P. Li, "Human-computer interaction system design based on surface EMG signals," in *Proceedings of 2014 International Conference on Modelling, Identification & Control*, Dec. 2014, pp. 94–98, doi: 10.1109/ICMIC.2014.7020734.

[56] M. C. Castro, S. P. Arjunan, and D. K. Kumar, "Selection of suitable hand gestures for reliable myoelectric human computer interface," *Biomed. Eng. Online*, vol. 14, no. 1, p. 30, Dec. 2015, doi: 10.1186/s12938-015-0025-5.

[57] G. E. Gopalakrishnan, "Identifying Hand Gestures Using sEMG For Human Machine Interaction," *ARPN J. Eng. Appl. Sci.*, vol. 11, pp. 12777–12785, 2016.

[58] N. M. Patil and S. R. Patil, "Review on real-time EMG acquisition and hand gesture recognition system," in *2017 International conference of Electronics, Communication and Aerospace Technology (ICECA)*, Apr. 2017, pp. 694–696, doi: 10.1109/ICECA.2017.8203629.

[59] S. P. Arjunan and D. K. Kumar, "Decoding subtle forearm flexions using fractal features of surface electromyogram from single and multiple sensors," *J. Neuroeng. Rehabil.*, vol. 7, no. 1, p. 53, Oct. 2010, doi: 10.1186/1743-0003-7-53.

[60] R. N. Khushaba, A. H. Al-Timemy, S. Kodagoda, and K. Nazarpour, "Combined influence of forearm orientation and muscular contraction on EMG pattern recognition," *Expert Syst. Appl.*, vol. 61, pp. 154–161, Nov. 2016, doi: 10.1016/J.ESWA.2016.05.031.

[61] P. Kaczmarek, T. Mánkowski, and J. Tomczýnski, "PutEMG—A surface electromyography hand gesture recognition dataset," *Sensors*, vol. 19, no. 16, p. 3548, Aug. 2019, doi: 10.3390/s19163548.

[62] C. Assad, M. Wolf, A. Stoica, T. Theodoridis, and K. Glette, "BioSleeve: A natural EMG-based interface for HRI," in *2013 8th ACM/IEEE International Conference on Human-Robot Interaction (HRI)*, Mar. 2013, pp. 69–70, doi: 10.1109/HRI.2013.6483505.

[63] P. Shenoy, K. J. Miller, B. Crawford, and R. P. N. Rao, "Online Electromyographic Control of a Robotic Prosthesis," *IEEE Trans. Biomed. Eng.*, vol. 55, no. 3, pp. 1128–1135, Mar. 2008, doi: 10.1109/TBME.2007.909536.

[64] C. DaSalla, J. Kim, and Y. Koike, "Robot control using electromyography (EMG) signals of the wrist," *Appl. Bionics Biomech.*, vol. 2, no. 2, pp. 97–102, Feb. 2005, doi: 10.1533/abbi.2004.0054.

[65] A. Wołczowski and M. Kurzyński, "Human-machine interface in bioprosthesis control using EMG signal classification," *Expert Syst.*, vol. 27, no. 1, pp. 53–70, Feb. 2010, doi: 10.1111/j.1468-0394.2009.00526.x.

[66] T. Lenzi, S. M. M. De Rossi, N. Vitiello, and M. C. Carrozza, "Intention-Based EMG Control for Powered Exoskeletons," *IEEE Trans. Biomed. Eng.*, vol. 59, no. 8, pp. 2180–2190, Aug. 2012, doi: 10.1109/TBME.2012.2198821.

[67] X. Chen, D. Zhang, and X. Zhu, "Application of a self-enhancing classification method to electromyography pattern recognition for multifunctional prosthesis control," *J. Neuroeng. Rehabil.*, vol. 10, no. 1, p. 44, May 2013, doi: 10.1186/1743-0003-10-44.

[68]  G. Purushothaman and K. K. Ray, "EMG based man-machine interaction - A pattern recognition research platform," *Rob. Auton. Syst.*, vol. 62, no. 6, pp. 864–870, Jun. 2014, doi: 10.1016/j.robot.2014.01.008.

[69]  M. T. Hammi, O. Salem, and A. Mehaoua, "An EMG-based Human-Machine Interface to control multimedia player," in *2015 17th International Conference on E-health Networking, Application & Services (HealthCom)*, Oct. 2015, pp. 274–279, doi: 10.1109/HealthCom.2015.7454511.

[70]  D. Farina, E. Fortunato, and R. Merletti, "Noninvasive estimation of motor unit conduction velocity distribution using linear electrode arrays," *IEEE Trans. Biomed. Eng.*, vol. 47, no. 3, pp. 380–388, Mar. 2000, doi: 10.1109/10.827303.

[71]  B. U. Kleine, J. H. Blok, R. Oostenveld, P. Praamstra, and D. F. Stegeman, "Magnetic stimulation-induced modulations of motor unit firings extracted from multi-channel surface EMG," *Muscle and Nerve*, vol. 23, no. 7, pp. 1005–1015, Jul. 2000, doi: 10.1002/1097-4598(200007)23:7<1005::AID-MUS2>3.0.CO;2-2.

[72]  S. M. Wood, J. A. Jarratt, A. T. Barker, and B. H. Brown, "Surface electromyography using electrode arrays: A study of motor neuron disease," *Muscle Nerve*, vol. 24, no. 2, pp. 223–230, Feb. 2001, doi: 10.1002/1097-4598(200102)24:2<223::AID-MUS70>3.0.CO;2-7.

[73]  B. U. Kleine, N. P. Schumann, D. F. Stegeman, and H. C. Scholle, "Surface EMG mapping of the human trapezius muscle: The topography of monopolar and bipolar surface EMG amplitude and spectrum parameters at varied forces and in fatigue," *Clin. Neurophysiol.*, vol. 111, no. 4, pp. 686–693, Apr. 2000, doi: 10.1016/S1388-2457(99)00306-5.

[74]  N. P. Schumann, H. C. Scholle, C. Anders, and E. Mey, "A topographical analysis of spectral electromyographic data of the human masseter muscle under different functional conditions in healthy subjects," *Arch. Oral Biol.*, vol. 39, no. 5, pp. 369–377, May 1994, doi: 10.1016/0003-9969(94)90166-X.

[75]  W. Geng, Y. Du, W. Jin, W. Wei, Y. Hu, and J. Li, "Gesture recognition by instantaneous surface EMG images.," *Sci. Rep.*, vol. 6, p. 36571, 2016, doi: 10.1038/srep36571.

[76]  C. Amma, T. Krings, J. Böer, and T. Schultz, "Advancing Muscle-Computer Interfaces with High-Density Electromyography," in *Proceedings of the 33rd Annual ACM Conference on Human Factors in Computing Systems*, 2015, pp. 929–938, doi: 10.1145/2702123.2702501.

[77]  G. Marco, B. Alberto, and V. Taian, "Surface EMG and muscle fatigue: multi-channel approaches to the study of myoelectric manifestations of muscle fatigue," *Physiol. Meas.*, vol. 38, no. 5, pp. R27–R60, May 2017, doi: 10.1088/1361-6579/aa60b9.

[78]  V. Becker, P. Oldrati, L. Barrios, and G. Sörös, "Touchsense: Classifying and measuring the force of finger touches with an electromyography armband," in *ACM International Conference Proceeding Series*, 2018, pp. 1–3, doi: 10.1145/3174910.3174947.

[79]  I. Donovan *et al.*, "MyoHMI: A low-cost and flexible platform for developing real-time human machine interface for myoelectric controlled applications," in *2016 IEEE International Conference on Systems, Man, and Cybernetics, SMC 2016 - Conference Proceedings*, Oct. 2017, pp. 4495–4500, doi: 10.1109/SMC.2016.7844940.

[80]  L. Chen, J. Fu, Y. Wu, H. Li, and B. Zheng, "Hand Gesture Recognition Using Compact CNN via Surface Electromyography Signals," *Sensors*, vol. 20, no. 3, p. 672, Jan. 2020, doi: 10.3390/s20030672.

[81]  X. Zhang, Z. Yang, T. Chen, D. Chen, and M. C. Huang, "Cooperative Sensing and Wearable Computing for Sequential Hand Gesture Recognition," *IEEE Sens. J.*, vol. 19, no. 14, pp. 5775–5783, Jul. 2019, doi: 10.1109/JSEN.2019.2904595.

[82]  J. O. Pinzón-Arenas, R. Jiménez-Moreno, and J. E. Herrera-Benavides, "Convolutional Neural Network for Hand Gesture Recognition using 8 different EMG Signals," in *2019 22nd Symposium on Image, Signal*

*Processing and Artificial Vision, STSIVA 2019 - Conference Proceedings*, Apr. 2019, pp. 1–5, doi: 10.1109/STSIVA.2019.8730272.

[83] C. Castellini, A. E. Fiorilla, and G. Sandini, "Multi-subject/daily-life activity EMG-based control of mechanical hands," *J. Neuroeng. Rehabil.*, vol. 6, no. 1, p. 41, Dec. 2009, doi: 10.1186/1743-0003-6-41.

[84] BTS Bioengineering, "FREEMG EMG Systems," 2019. .

[85] A. Wege and A. Zimmermann, "Electromyography Sensor Based Control for a Hand Exoskeleton," pp. 1470–1475, 2008.

[86] M. Simão, N. Mendes, and O. Gibaru, "A Review on Electromyography Decoding and Pattern Recognition for Human-Machine Interaction," vol. 7, 2020, doi: 10.1109/ACCESS.2019.2906584.

[87] J. S. Chang, A. F. Facchetti, and R. Reuss, "A Circuits and Systems Perspective of Organic/Printed Electronics: Review, Challenges, and Contemporary and Emerging Design Approaches," *IEEE J. Emerg. Sel. Top. Circuits Syst.*, vol. 7, no. 1, pp. 7–26, Mar. 2017, doi: 10.1109/JETCAS.2017.2673863.

[88] A. Moin *et al.*, "A wearable biosensing system with in-sensor adaptive machine learning for hand gesture recognition," *Nat. Electron.*, vol. (in press), 2021, doi: 10.1038/s41928-020-00510-8.

[89] X. Chen, X. Zhang, Z.-Y. Zhao, J. Yang, V. Lantz, and K. Wang, "Hand Gesture Recognition Research Based on Surface EMG Sensors and 2D-accelerometers," in *2007 11th IEEE International Symposium on Wearable Computers*, Oct. 2007, pp. 1–4, doi: 10.1109/ISWC.2007.4373769.

[90] J. Kim, J. Wagner, M. Rehm, and E. André, "Bi-channel sensor fusion for automatic sign language recognition," in *2008 8th IEEE International Conference on Automatic Face & Gesture Recognition*, Sep. 2008, pp. 1–6, doi: 10.1109/AFGR.2008.4813341.

[91] X. Zhang, X. Chen, Y. Li, V. Lantz, K. Wang, and J. Yang, "A Framework for Hand Gesture Recognition Based on Accelerometer and EMG Sensors," *IEEE Trans. Syst. Man, Cybern. - Part A Syst. Humans*, vol. 41, no. 6, pp. 1064–1076, Nov. 2011, doi: 10.1109/TSMCA.2011.2116004.

[92] A. Fougner, E. Scheme, A. D. C. Chan, K. Englehart, and Ø. Stavdahl, "Resolving the Limb Position Effect in Myoelectric Pattern Recognition," *IEEE Trans. Neural Syst. Rehabil. Eng.*, vol. 19, no. 6, pp. 644–651, Dec. 2011, doi: 10.1109/TNSRE.2011.2163529.

[93] A. Fougner, E. Scheme, A. D. C. Chan, K. Englehart, and Ø. Stavdahl, "A multi-modal approach for hand motion classification using surface EMG and accelerometers," in *2011 Annual International Conference of the IEEE Engineering in Medicine and Biology Society*, Aug. 2011, pp. 4247–4250, doi: 10.1109/IEMBS.2011.6091054.

[94] A. Gijsberts, M. Atzori, C. Castellini, H. Müller, and B. Caputo, "Movement Error Rate for Evaluation of Machine Learning Methods for sEMG-Based Hand Movement Classification," *IEEE Trans. NEURAL Syst. Rehabil. Eng.*, vol. 22, no. 4, p. 735, 2014, doi: 10.1109/TNSRE.2014.2303394.

[95] J. Wu, L. Sun, and R. Jafari, "A Wearable System for Recognizing American Sign Language in Real-Time Using IMU and Surface EMG Sensors," *IEEE J. Biomed. Heal. Informatics*, vol. 20, no. 5, pp. 1281–1290, Sep. 2016, doi: 10.1109/JBHI.2016.2598302.

[96] V. Dhakal, A. M. Feit, P. O. Kristensson, and A. Oulasvirta, "Observations on typing from 136 million keystrokes," in *Conference on Human Factors in Computing Systems - Proceedings*, 2018, vol. 2018-April, pp. 1–12, doi: 10.1145/3173574.3174220.

[97] O. Vinyals *et al.*, "StarCraft II: A New Challenge for Reinforcement Learning," Aug. 2017.

[98] U. Bellugi, S. Fischer, and C. Newkirk, "The rate of speaking and signing," in *The Signs of Language*, 1979, pp. 181–194.

[99] R. B. Wilbur, "Effects of varying rate of signing on ASL manual signs and nonmanual markers.," *Lang. Speech*, vol. 52, no. Pt 2-3, pp. 245–85, 2009, doi: 10.1177/0023830909103174.

[100] S. K. Liddell and R. E. Johnson, "American Sign Language: The Phonological Base," *Sign Lang. Stud.*, vol. 1064, no. 1, pp. 195–277, 1989, doi: 10.1353/sls.1989.0027.

[101] W. Sandler, "The Phonological Organization of Sign Languages," *Linguist. Lang. Compass*, vol. 6, no. 3, pp. 162–182, Mar. 2012, doi: 10.1002/lnc3.326.

[102] P. M. Fitts, "The information capacity of the human motor system in controlling the amplitude of movement," *J. Exp. Psychol.*, vol. 47, no. 6, pp. 381–391, 1954, doi: 10.1037/h0055392.

[103] I. S. MacKenzie, *Human-Computer Interaction: An Empirical Research Perspective*. Elsevier Science, 2013.

[104] R. A. Burno, B. Wu, R. Doherty, H. Colett, and R. Elnaggar, "Applying Fitts' Law to Gesture Based Computer Interactions," *Procedia Manuf.*, vol. 3, pp. 4342–4349, Jan. 2015, doi: 10.1016/j.promfg.2015.07.429.

[105] L. Sambrooks and B. Wilkinson, "Comparison of gestural, touch, and mouse interaction with Fitts' law," in *Proceedings of the 25th Australian Computer-Human Interaction Conference: Augmentation, Application, Innovation, Collaboration, OzCHI 2013*, 2013, pp. 119–122, doi: 10.1145/2541016.2541066.

[106] G. D. Langolf, D. B. Chaffin, and J. A. Foulke, "An Investigation of Fitts' Law Using a Wide Range of Movement Amplitudes," *J. Mot. Behav.*, vol. 8, no. 2, pp. 113–128, Jun. 1976, doi: 10.1080/00222895.1976.10735061.

[107] Y. Wu, B. Zheng, and Y. Zhao, "Dynamic Gesture Recognition Based on LSTM-CNN," in *2018 Chinese Automation Congress (CAC)*, Nov. 2018, pp. 2446–2450, doi: 10.1109/CAC.2018.8623035.

[108] H. Chen, Y. Zhang, Z. Zhang, Y. Fang, H. Liu, and C. Yao, "Exploring the relation between EMG sampling frequency and hand motion recognition accuracy," in *2017 IEEE International Conference on Systems, Man, and Cybernetics (SMC)*, Oct. 2017, pp. 1139–1144, doi: 10.1109/SMC.2017.8122765.

[109] G. Li, Y. Li, L. Yu, and Y. Geng, "Conditioning and Sampling Issues of EMG Signals in Motion Recognition of Multifunctional Myoelectric Prostheses," *Ann. Biomed. Eng.*, vol. 39, no. 6, pp. 1779–1787, Jun. 2011, doi: 10.1007/s10439-011-0265-x.

[110] Y. Li, X. Chen, X. Zhang, and P. Zhou, "Several practical issues toward implementing myoelectric pattern recognition for stroke rehabilitation," *Med. Eng. Phys.*, vol. 36, no. 6, pp. 754–760, 2014, doi: 10.1016/j.medengphy.2014.01.005.

[111] B. Hudgins, P. A. Parker, and R. N. Scott, "A New Strategy for Multifunction Myoelectric Control," *IEEE Trans. Biomed. Eng.*, vol. 40, no. 1, pp. 82–94, 1993, doi: 10.1109/10.204774.

[112] K. Englehart, B. Hudgins, and P. A. Parker, "A wavelet-based continuous classification scheme for multifunction myoelectric control," *IEEE Trans. Biomed. Eng.*, vol. 48, no. 3, pp. 302–311, 2001, doi: 10.1109/10.914793.

[113] M. Rojas-Martínez, M. A. Mañanas, J. F. Alonso, and R. Merletti, "Identification of isometric contractions based on High Density EMG maps," *J. Electromyogr. Kinesiol.*, vol. 23, no. 1, pp. 33–42, Feb. 2013, doi: 10.1016/J.JELEKIN.2012.06.009.

[114] K. Xing *et al.*, "Hand Gesture Recognition Based on Deep Learning Method," in *2018 IEEE Third International Conference on Data Science in Cyberspace (DSC)*, Jun. 2018, pp. 542–546, doi: 10.1109/DSC.2018.00087.

[115] K. Englehart and B. Hudgins, "A robust, real-time control scheme for multifunction myoelectric control," *IEEE Trans. Biomed. Eng.*, vol. 50, no. 7, pp. 848–854, Jul. 2003, doi: 10.1109/TBME.2003.813539.

[116] C. Castellini and P. van der Smagt, "Surface EMG in advanced hand prosthetics," *Biol. Cybern.*, vol. 100, no. 1, pp. 35–47, Jan. 2009, doi: 10.1007/s00422-008-0278-1.

[117] U. Côté-Allard *et al.*, "Deep Learning for Electromyographic Hand Gesture Signal Classification Using Transfer Learning," *IEEE Trans. Neural Syst. Rehabil. Eng.*, vol. 27, no. 4, pp. 760–771, Apr. 2019, doi: 10.1109/TNSRE.2019.2896269.

[118] Y. Du, W. Jin, W. Wei, Y. Hu, and W. Geng, "Surface EMG-based inter-session gesture recognition enhanced by deep domain adaptation," *Sensors*, vol. 17, no. 3, p. 458, 2017.

[119] U. Côté-Allard, C. L. Fall, A. Campeau-Lecours, C. Gosselin, F. Laviolette, and B. Gosselin, "Transfer learning for sEMG hand gestures recognition using convolutional neural networks," in *2017 IEEE International Conference on Systems, Man, and Cybernetics (SMC)*, Oct. 2017, pp. 1663–1668, doi: 10.1109/SMC.2017.8122854.

[120] M. A. Oskoei and H. Hu, "Myoelectric control systems—A survey," *Biomed. Signal Process. Control*, vol. 2, no. 4, pp. 275–294, Oct. 2007, doi: 10.1016/J.BSPC.2007.07.009.

[121] A. Phinyomark, P. Phukpattaranont, and C. Limsakul, "Feature reduction and selection for EMG signal classification," *Expert Syst. Appl.*, vol. 39, no. 8, pp. 7420–7431, Jun. 2012, doi: 10.1016/J.ESWA.2012.01.102.

[122] A. Phinyomark, F. Quaine, S. Charbonnier, C. Serviere, F. Tarpin-Bernard, and Y. Laurillau, "EMG feature evaluation for improving myoelectric pattern recognition robustness," *Expert Syst. Appl.*, vol. 40, no. 12, pp. 4832–4840, Sep. 2013, doi: 10.1016/J.ESWA.2013.02.023.

[123] K. Englehart, B. Hudgins, P. A. Parker, and M. Stevenson, "Classification of the myoelectric signal using time-frequency based representations," *Med. Eng. Phys.*, vol. 21, no. 6–7, pp. 431–438, Jul. 1999, doi: 10.1016/S1350-4533(99)00066-1.

[124] M. Zardoshti-Kermani, B. C. Wheeler, K. Badie, and R. M. Hashemi, "EMG feature evaluation for movement control of upper extremity prostheses," *IEEE Trans. Rehabil. Eng.*, vol. 3, no. 4, pp. 324–333, 1995, doi: 10.1109/86.481972.

[125] R. Boostani and M. H. Moradi, "Evaluation of the forearm EMG signal features for the control of a prosthetic hand," *Physiol. Meas.*, vol. 24, no. 2, pp. 309–319, May 2003, doi: 10.1088/0967-3334/24/2/307.

[126] J. Qi, G. Jiang, G. Li, Y. Sun, and B. Tao, "Intelligent human-computer interaction based on surface EMG gesture recognition," *IEEE Access*, vol. 7, pp. 61378–61387, 2019.

[127] A. Doswald, F. Carrino, and F. Ringeval, "Advanced processing of sEMG signals for user independent gesture recognition," in *XIII Mediterranean Conference on Medical and Biological Engineering and Computing 2013*, 2013, pp. 758–761.

[128] A. Phinyomark, R. N. Khushaba, and E. Scheme, "Feature Extraction and Selection for Myoelectric Control Based on Wearable EMG Sensors.," *Sensors (Basel).*, vol. 18, no. 5, May 2018, doi: 10.3390/s18051615.

[129] R. N. Khushaba and S. Kodagoda, "Electromyogram (EMG) feature reduction using Mutual Components Analysis for multifunction prosthetic fingers control," in *2012 12th International Conference on Control Automation Robotics & Vision (ICARCV)*, Dec. 2012, pp. 1534–1539, doi: 10.1109/ICARCV.2012.6485374.

[130] M. Atzori *et al.*, "Characterization of a Benchmark Database for Myoelectric Movement Classification," *IEEE Trans. Neural Syst. Rehabil. Eng.*, vol. 23, no. 1, pp. 73–83, Jan. 2015, doi: 10.1109/TNSRE.2014.2328495.

[131] W. Wei, Y. Wong, Y. Du, Y. Hu, M. Kankanhalli, and W. Geng, "A multi-stream convolutional neural network for sEMG-based gesture recognition in muscle-computer interface," *Pattern Recognit. Lett.*, vol. 119, pp. 131–138, Mar. 2019, doi: 10.1016/J.PATREC.2017.12.005.

[132] Y. Huang, K. Englehart, B. Hudgins, and A. D. C. Chan, "A Gaussian Mixture Model Based Classification Scheme for Myoelectric Control of Powered Upper Limb Prostheses," *IEEE Trans. Biomed. Eng.*, vol. 52, no. 11, pp. 1801–1811, Nov. 2005, doi: 10.1109/TBME.2005.856295.

[133] U. Côté-Allard, E. Campbell, A. Phinyomark, F. Laviolette, B. Gosselin, and E. Scheme, "Interpreting Deep Learning Features for Myoelectric Control: A Comparison With Handcrafted Features," *Front. Bioeng. Biotechnol.*, vol. 8, p. 158, Mar. 2020, doi: 10.3389/fbioe.2020.00158.

[134] A. Krizhevsky, I. Sutskever, and G. E. Hinton, "ImageNet classification with deep convolutional neural networks," *Commun. ACM*, vol. 60, no. 6, pp. 84–90, May 2017, doi: 10.1145/3065386.

[135] C. Szegedy, S. Ioffe, V. Vanhoucke, and A. Alemi, "Inception-v4, Inception-ResNet and the Impact of Residual Connections on Learning," Feb. 2016.

[136] F. A. Gers, J. Schmidhuber, and F. Cummins, "Learning to forget: Continual prediction with LSTM," in *IEE Conference Publication*, 1999, vol. 2, no. 470, pp. 850–855, doi: 10.1049/cp:19991218.

[137] R. N. Khushaba, S. Kodagoda, M. Takruri, and G. Dissanayake, "Toward improved control of prosthetic fingers using surface electromyogram (EMG) signals," *Expert Syst. Appl.*, vol. 39, no. 12, pp. 10731–10738, Sep. 2012, doi: 10.1016/J.ESWA.2012.02.192.

[138] M. Atzori, M. Cognolato, and H. Müller, "Deep Learning with Convolutional Neural Networks Applied to Electromyography Data: A Resource for the Classification of Movements for Prosthetic Hands," *Front. Neurorobot.*, vol. 10, Sep. 2016, doi: 10.3389/fnbot.2016.00009.

[139] P. Tsinganos, B. Cornelis, J. Cornelis, B. Jansen, and A. Skodras, "Deep learning in EMG-based gesture recognition," *PhyCS 2018 - Proc. 5th Int. Conf. Physiol. Comput. Syst.*, pp. 107–114, 2018, doi: 10.5220/0006960201070114.

[140] W. Geng, Y. Hu, Y. Wong, W. Wei, Y. Du, and M. Kankanhalli, "A novel attention-based hybrid CNN-RNN architecture for sEMG-based gesture recognition," *PLoS One*, vol. 13, no. 10, pp. 1–18, 2018, doi: 10.1371/journal.pone.0206049.

[141] J. Chen, S. Bi, G. Zhang, and G. Cao, "High-density surface emg-based gesture recognition using a 3d convolutional neural network," *Sensors*, vol. 20, no. 4, p. 1201, Feb. 2020, doi: 10.3390/s20041201.

[142] N. Nahid, A. Rahman, and M. A. R. Ahad, "Deep Learning Based Surface EMG Hand Gesture Classification for Low-Cost Myoelectric Prosthetic Hand," 2020.

[143] T. Bao, S. A. R. Zaidi, S. Xie, P. Yang, and Z.-Q. Zhang, "A CNN-LSTM Hybrid Model for Wrist Kinematics Estimation Using Surface Electromyography," *IEEE Transactions on Instrumentation and Measurement*. pp. 1–1, 2020, doi: 10.1109/TIM.2020.3036654.

[144] Q. Zhang, D. Wang, R. Zhao, and Y. Yu, "MyoSign: Enabling End-to-End Sign Language Recognition with Wearables," in *Proceedings of the 24th International Conference on Intelligent User Interfaces*, Mar. 2019, pp. 650–660, doi: 10.1145/3301275.3302296.

[145] P. Xia, J. Hu, and Y. Peng, "EMG-Based Estimation of Limb Movement Using Deep Learning With Recurrent Convolutional Neural Networks," *Artif. Organs*, vol. 42, no. 5, pp. E67–E77, May 2018, doi: 10.1111/aor.13004.

[146] M. Atzori, A. Gijsberts, H. Muller, and B. Caputo, "Classification of hand movements in amputated subjects by sEMG and accelerometers," in *2014 36th Annual International Conference of the IEEE Engineering in Medicine and Biology Society*, Aug. 2014, pp. 3545–3549, doi: 10.1109/EMBC.2014.6944388.

[147] R. N. Khushaba, S. Kodagoda, D. Liu, and G. Dissanayake, "Muscle computer interfaces for driver distraction reduction.," *Comput. Methods Programs Biomed.*, vol. 110, no. 2, pp. 137–49, May 2013, doi: 10.1016/j.cmpb.2012.11.002.

[148] A. H. Al-Timemy, R. N. Khushaba, G. Bugmann, and J. Escudero, "Improving the Performance Against Force Variation of EMG Controlled Multifunctional Upper-Limb Prostheses for Transradial Amputees," *IEEE Trans. Neural Syst. Rehabil. Eng.*, vol. 24, no. 6, pp. 650–661, Jun. 2016, doi: 10.1109/TNSRE.2015.2445634.

[149] R. N. Khushaba, M. Takruri, J. V. Miro, and S. Kodagoda, "Towards limb position invariant myoelectric pattern recognition using time-dependent spectral features," *Neural Networks*, vol. 55, pp. 42–58, Jul. 2014, doi: 10.1016/J.NEUNET.2014.03.010.

[150] M. Atzori *et al.*, "Building the Ninapro database: A resource for the biorobotics community," *Proc. IEEE RAS EMBS Int. Conf. Biomed. Robot. Biomechatronics*, no. Section II, pp. 1258–1265, 2012, doi: 10.1109/BioRob.2012.6290287.

[151] M. Atzori *et al.*, "Electromyography data for non-invasive naturally-controlled robotic hand prostheses," *Sci. data*, vol. 1, no. 1, pp. 1–13, 2014.

[152] S. Pizzolato, L. Tagliapietra, M. Cognolato, M. Reggiani, H. Müller, and M. Atzori, "Comparison of six electromyography acquisition setups on hand movement classification tasks," *PLoS One*, vol. 12, no. 10, p. e0186132, Oct. 2017, doi: 10.1371/journal.pone.0186132.

[153] W. Jin, Y. Li, and S. Lin, "Design of a novel non-invasive wearable device for array surface electromyogram," *Int. J. Inf. Electron. Eng.*, vol. 6, no. 2, p. 139, 2016.